\documentclass[aps,prd,showpacs,notitlepage,twocolumn,superscriptaddress,nofootinbib,preprintnumbers]{revtex4-2}
\usepackage{wrapfig}
\usepackage{bbm}
\usepackage{url}
\usepackage{mathrsfs}
\usepackage{epsfig}
\usepackage[utf8]{inputenc}
\usepackage{soul,xcolor}
\usepackage{graphicx}
\usepackage{amsfonts}
\usepackage{amsthm}
\usepackage[figuresright]{rotating}
\usepackage{amssymb}
\usepackage{amsmath}
\usepackage{dcolumn}
\usepackage{physics}
\usepackage{float}
\usepackage{bm}
\usepackage{physics}
\usepackage{verbatim}
\usepackage{braket}
\usepackage[normalem]{ulem}
\usepackage[ruled,vlined,linesnumbered]{algorithm2e}
\usepackage{setspace}
\usepackage{lipsum}
\usepackage[colorlinks,linkcolor=blue,anchorcolor=blue,citecolor=blue,urlcolor=blue]{hyperref}
\usepackage{graphicx}
\usepackage{amsmath, nccmath}
\usepackage{amsfonts}
\usepackage{amssymb}
\usepackage{esvect}
\usepackage{amsthm}
\usepackage{mathdots}
\usepackage[caption=false]{subfig}
\usepackage{mathtools}
\usepackage{tensor}
\usepackage{mathrsfs}
\usepackage{comment}
\usepackage{bbm}
\usepackage{graphicx}
\usepackage{ragged2e}
\usepackage{color}
\usepackage{wrapfig}
\usepackage{color}
\usepackage{graphicx}
\usepackage{tikz}
\usepackage{dsfont}
\usepackage{mathbbol}
\usepackage{amssymb} 
\usepackage{simpler-wick}

\DeclareSymbolFontAlphabet{\amsmathbb}{AMSb}
\usetikzlibrary{shapes.misc}
\tikzset{cross/.style={cross out, draw=black, minimum size=2*(#1-\pgflinewidth), inner sep=0pt, outer sep=0pt},
cross/.default={3pt}}
\usetikzlibrary{arrows,shapes,positioning}
\usetikzlibrary{decorations.markings}
\usepackage[rightcaption]{sidecap}
\tikzstyle arrowstyle=[scale=1]
\tikzstyle directed=[postaction={decorate,decoration={markings,
    mark=at position .65 with {\arrow[arrowstyle]{stealth}}}}]
\tikzstyle reverse directed=[postaction={decorate,decoration={markings,
    mark=at position .65 with {\arrowreversed[arrowstyle]{stealth};}}}]
\usetikzlibrary{positioning}

\newcommand{\bea}{\begin{eqnarray}}
\newcommand{\eea}{\end{eqnarray}}
\newcommand{\be}{\begin{eqnarray}}
\newcommand{\ee}{\end{eqnarray}}
\newcommand{\bma}{\begin{matrix}}
\newcommand{\ema}{\cr\end{matrix}}

\newcommand{\SU}{ {\text {SU}}}

\newcommand{\U}{ {\text {U}}}

\newcommand{\complex}{ {\amsmathbb C} }
\newcommand{\integers}{ {\amsmathbb Z} }

\usepackage{ragged2e}
\usepackage{etoolbox}   
\usepackage{amsmath, amssymb}

\usepackage{hyperref}

\usepackage{ragged2e}
\usepackage{etoolbox}
\makeatletter
\patchcmd{\@makecaption}
  {\centering}
  {\justifying}
  {}
  {}
\makeatother

\def\cB{{\cal B}}

\def\cG{{\cal G}}
\def\cH{{\cal H}}

\def\cK{{\cal K}}

\def\cM{{\cal M}}
\def\cN{{\cal N}}
\def\cO{{\cal O}}

\def\cQ{{\cal Q}}

\def\cT{{\cal T}}
\def\cU{{\cal U}}
\def\cV{{\cal V}}
\def\cW{{\cal W}}

\def\Tr{{\rm Tr}}

\def\half{{1\over 2}}

\def\p{\partial}

\def\a{\alpha}
\def\b{\beta}

\def\d{\delta}

\def\th{{\rm th}}

\def\sfU{{\sf U}}

\def\sfE{{\sf E}}
\def\sfB{{\sf B}}
\def\sfG{{\sf G}}

\def\no{\nonumber}

\definecolor{darkred}{rgb}{0.8,0.1,0.1}
\hypersetup{colorlinks=true, linkcolor=darkred, citecolor=blue, linktoc=page}

\newcommand{\nc}{\newcommand}
\nc{\rnc}{\renewcommand} 

\rnc{\a}{\alpha}
\rnc{\b}{\beta}
\rnc{\d}{\delta}
\nc{\e}{\epsilon}
\nc{\z}{\zeta}
\nc{\m}{\mu}
\nc{\n}{\nu}
\rnc{\r}{\rho}
\rnc{\k}{\kappa}
\rnc{\l}{\lambda}
\nc{\s}{\sigma}
\rnc{\t}{\tau}
\nc{\w}{\omega}
\nc{\x}{\chi}
\nc{\F}{\Phi}
\rnc{\L}{\Lambda}

\nc{\pd}{\partial}

\makeatletter
\newcommand{\vast}{\bBigg@{4}}
\newcommand{\Vast}{\bBigg@{5}}
\makeatother

\begin{document}

\title{Hamiltonian Lattice QED$_3$ with One and Two Flavors of Wilson Fermions:\\
Topological Structure and Response}

\author{Sriram Bharadwaj}
\affiliation{
Mani L. Bhaumik Institute for Theoretical Physics, Department of Physics and Astronomy,\\ University of California, Los Angeles, CA 90095, USA
}
\author{Emil Rosanowski}
\affiliation{Transdisciplinary Research Area ``Building Blocks of Matter and Fundamental Interactions'' (TRA Matter) and Helmholtz Institute for Radiation and Nuclear Physics (HISKP), University of Bonn, Nussallee 14-16, 53115 Bonn, Germany}
\author{Simran Singh}
\affiliation{Transdisciplinary Research Area ``Building Blocks of Matter and Fundamental Interactions'' (TRA Matter) and Helmholtz Institute for Radiation and Nuclear Physics (HISKP), University of Bonn, Nussallee 14-16, 53115 Bonn, Germany}
\author{Alice Di Tucci}
\affiliation{Deutsches Elektronen-Synchrotron DESY, Platanenallee 6, 15738 Zeuthen, Germany}
\author{Changnan Peng}
\affiliation{Department of Physics, Massachusetts Institute of Technology, Cambridge, Massachusetts 02139, USA}
\author{Karl Jansen}
\affiliation{Deutsches Elektronen-Synchrotron DESY, Platanenallee 6, 15738 Zeuthen, Germany}
\affiliation{
 Computation-Based Science and Technology Research Center, The Cyprus Institute, 20 Kavafi Street,
2121 Nicosia, Cyprus
}
\author{Lena Funcke}
\affiliation{Transdisciplinary Research Area ``Building Blocks of Matter and Fundamental Interactions'' (TRA Matter) and Helmholtz Institute for Radiation and Nuclear Physics (HISKP), University of Bonn, Nussallee 14-16, 53115 Bonn, Germany}
\author{Di Luo}
\thanks{diluo1000@gmail.com}
\affiliation{Department of Physics, Tsinghua University, Beijing 100084, China}
\affiliation{Institute of Advanced Study, Tsinghua University, Beijing 100084, China}
\affiliation{Department of Electrical and Computer Engineering, University of California, Los Angeles, CA 90095, USA}

\begin{abstract}
The quantum simulation of topological phases in (2+1)D quantum electrodynamics with Wilson fermions provides a promising route toward realizing topological phenomena in near-term lattice experiments. We show that the commonly used staggered-fermion discretizations in Hamiltonian gauge theories possesses an exact time-reversal symmetry, which forbids the emergence of nontrivial topological phases and has led to confusion in the existing literature. In this work, we resolve this obstacle by systematically analyzing fermion discretization effects in (2+1)D lattice Hamiltonians of fermions coupled to $\mathrm{U}(1)$ gauge fields that satisfy Gauss’ law. We show that Wilson fermions, already in the minimal one-flavor theory, naturally enable topological regimes with nonzero Chern numbers, and that the two-flavor extension at finite chemical potential further enriches the accessible topological structure. We develop gauge-invariant diagnostics of topological response, including many-body Chern numbers and current correlators that remain robust probes at weak coupling. Finally, through extensive exact diagonalization calculations across both flavor settings, we characterize the spectrum, correlators, and topological invariants, providing a concrete foundation for near-term quantum simulations of topological phases in lattice field theories. The implications of this work for quantum simulations of lattice field theory are analyzed in a joint submission \cite{Bharadwaj:2025idp}. 
\end{abstract}
\maketitle

\section{Introduction} 
Quantum simulation is rapidly becoming a leading paradigm for accessing nonperturbative regimes of quantum field theories, offering a controlled route to strongly correlated dynamics that remain inaccessible to classical computation. In this context, (2+1)D quantum electrodynamics (QED$_3$) occupies a special position: it provides one of the simplest interacting gauge theories with confinement and chiral phenomena, while also supporting the possibility of intrinsically topological gauge responses relevant to condensed-matter realizations. Recent progress in quantum hardware, together with advances in tensor-network and neural-network representations of gauge-invariant states, has made Hamiltonian lattice formulations of QED$_3$ a particularly promising target for near-term experiments. A central remaining difficulty, however, is to construct lattice gauge models that simultaneously respect Gauss’ law, faithfully encode fermionic dynamics, and capture the emergence of topological structure beyond perturbative continuum arguments. Establishing such a framework is essential for turning QED$_3$ into a practical platform for quantum simulation of interacting topological phases.

An important theoretical challenge in Hamiltonian lattice QED$_3$ is to understand how topological phenomena depend on the choice of fermion regularization once Gauss’ law is imposed microscopically. Although staggered fermions are frequently adopted in quantum simulation proposals, their capability to realize lattice topological phases has remained unsettled, giving rise to contradictory claims regarding Chern numbers and Chern–Simons physics in $(2+1)$ dimensions. Here we show that this ambiguity has a simple origin: staggered fermions in the Hamiltonian gauge-invariant setting exhibit an exact time-reversal constraint that excludes nontrivial topological phases. We then demonstrate that Wilson fermions provide a natural route toward topological lattice gauge matter, and that both the minimal one-flavor formulation and its two-flavor generalization at finite chemical potential support a substantially richer topological landscape. This enables a unified characterization of the resulting topological structure through gauge-invariant invariants and response functions, including many-body Chern numbers and current-based correlators. Together with extensive exact diagonalization calculations in both flavor sectors, our analysis establishes Wilson-fermion QED$_3$ as a concrete and experimentally relevant framework for accessing topological response in quantum simulations of gauge theories.

In this work\footnote{This manuscript provides the detailed analysis and additional results for a joint submission; see Ref.~\cite{Bharadwaj:2025idp} for a complementary presentation.}, we systematically explore the role of fermion discretization in the emergence of topological phases in (2+1)D lattice Hamiltonians of fermions coupled to $\U(1)$ gauge fields that satisfy Gauss' law. First, we demonstrate that staggered fermions fail to support nontrivial topological phases, providing a resolution to the existing confusion in the literature. Second, we show that Wilson fermions, with both one- and two-species in the presence of finite chemical potential, naturally support a variety of topological phases, including Chern insulator and quantum spin Hall with nonzero Chern numbers in the weak coupling limit. This analysis provides a theoretical foundation for constructing lattice gauge models that accurately encode topological effects in quantum simulations. Furthermore, we analyze current correlators, and show that they are robust probes for the existence of topological phases at weak coupling. Finally, we carry out a detailed numerical study for both the one- and two- flavor cases, including the spectrum, current-correlators, and the many-body Chern number via exact diagonalization. Our results offer a crucial step toward realizing near-term quantum simulations of topological phases of gauge theories.

This paper is organized as follows. In Sec.~\ref{sec:TimeReversal} we establish the time-reversal symmetry of staggered fermions in $(2+1)$D Hamiltonian gauge theories, which forbids the emergence of topological phases. In Sec.~\ref{sec:Nf=1} we introduce the one-flavor Wilson-fermion Hamiltonian coupled to $\U(1)$ gauge fields and characterize its topological phases via the Chern number. In Sec.~\ref{sec:Nf=2} we extend the analysis to two flavors at finite chemical potential and show that this theory hosts both Integer Quantum Hall (IQH) and Quantum Spin Hall (QSH) effects. In Sec.~\ref{sec:mbchern_robust} we define and compute the many-body Chern number in the gauge-invariant Hilbert space and analyze its perturbative robustness at finite gauge coupling. In Sec.~\ref{sec:CurrentCorr} we develop gauge-invariant current correlators as probes of topological response. In Sec.~\ref{sec:Numerics} we present our exact diagonalization results for spectra, current correlators, and the many-body Chern number. In Sec.~\ref{sec:TruncationFiniteSize}, we provide a systematic analysis of truncation and finite-size effects. In Sec.~\ref{sec:Conclusion}, we discuss the implications of our work for near-term quantum simulation.

\section{Time-Reversal Symmetry and Topological Phases} \label{sec:TimeReversal}

In this section, we explain the relation between time-reversal symmetry and topological phases characterized by a Chern number and highlight an apparent conflict between the continuum and the lattice. 

In the continuum, the theory of a single Dirac fermion coupled to a $\U(1)$ gauge field exhibits the so-called parity ``anomaly",\footnote{This is \textit{not} an anomaly, but rather an explicit breaking. This terminology is simply a convention (see Chap. 8 of~\cite{TongGaugeTheory2018}).} which is manifest in its effective low-energy Chern-Simons description. In other words, since charge conjugation is a UV symmetry and CPT is always a symmetry, both time-reversal and parity are broken explicitly in the continuum. In contrast, we show below that the lattice Hamiltonian formulation of a single staggered fermion coupled to a $\U(1)$ gauge field is time-reversal invariant, implying that the Chern number of the Berry curvature vanishes. Therefore, the low-energy limit of staggered fermions is always in a topologically trivial phase with a vanishing Chern-Simons level.

Consider a theory of $(2+1)$D staggered one-component complex fermions $\chi_r$ coupled to $\U(1)$ gauge fields $\sfU_k(r)$ with $k=x, y$. We will always treat time as a continuum variable with space-time signature $(+,-,-)$ and space as an $L\times L$ square lattice with periodic boundary conditions and unit lattice-spacing. The fermionic Hamiltonian is:
\begin{align}
    H_s= \sum_r \Big[\frac{i}{2}\chi_{r+\hat{x}}^\dagger\sfU_{x}(r)\chi_r-\frac{(-)^{i+j}}{2}\chi_{r+\hat{y}}^\dagger\sfU_{y}(r)\chi_r\Big] + h.c.\no\\+ m\sum_{r}(-)^{i+j}\chi_r^\dagger \chi_r
\end{align}
where $r=(i,j)$ labels the sites. The local charge-density operator is defined as $\cQ^s_r = \chi_r^\dagger \chi_r - \frac{1-(-)^{i+j}}{2}$. Since we are analyzing the IR physics, we seek an effective Hamiltonian that correctly models the ground-state of the theory including the symmetries. In particular, the ground state must satisfy lattice-translational symmetry, as required by the continuum limit. Exact diagonalization of small systems (e.g., $L=4$) shows the existence of the following staggered charge pattern $\cQ^s_{i,j} = \cQ^s_{i+1,j+1},\cQ^s_{i,j+1} = \cQ^s_{i+1,j}$ for any site $r = (i,j)$ provided the gauge fields on the links satisfy $\sfU_k(r) = \sfU_k({r+h}) =\sfU_k({r+d})$, where $h = (2,0)$, $d=(1,1)$. Hence, we take the unit cell to be spanned by the vectors $\{h, d\}$ \cite{chen2022simulating}.\footnote{Note that this is the same pattern reported in \cite{chen2022simulating}. The main difference here is that we also have a staggered hopping term, while, in \cite{chen2022simulating}, only the mass term is staggered.} In this language, the Hamiltonian is
\begin{align}
    H_m &= m \sum_{r} [\psi_{a,r}^\dagger \psi_{a,r} - \psi_{b,r}^\dagger\psi_{b,r}]\\
    H_\text{hop}&= \frac{1}{2} \sum_{r}\Big[ie^{i\phi_1}\psi_{b,r}^\dagger\psi_{a,r}+ie^{-i\phi_3}\psi_{b,r-h}^\dagger\psi_{a,r}\no\\&-e^{i\phi_4}\psi_{b,r-h+d}^\dagger\psi_{a,r}+e^{-i\phi_2}\psi_{b,r-d}^\dagger\psi_{a,r}\Big]+h.c.
\end{align}
where we call the sites $a$ if $(i+j)$ is even or $b$ when $(i+j)$ is odd. In Fourier space, the total Hamiltonian is
\begin{align}
    H &= \sum_k b_k^\dagger F a_k +h.c. +m\sum_k \left[a_k^\dagger a_k - b_k^\dagger b_k \right]
\end{align}
where $F=\half(ie^{i\phi_1}+ ie^{-i\phi_3+2ik_x}-e^{i\phi_4+ik_x-ik_y}+e^{-i\phi_2+ik_x+ik_y})$
and 
\begin{align}
    \psi_{r,a} &= \frac{1}{L}\sum_{k}e^{ik\vdot r}a_k &\psi_{r,b} &= \frac{1}{L}\sum_{k}e^{ik\vdot r}b_k
\end{align}
Decomposing into real and imaginary parts $F = \cH^s_x + i\cH_y^s$ with $\cH_z^s = m$, this may trivially be reorganized into \eqref{StagH}. Hence, the low-energy Hamiltonian is
\begin{align}\label{StagH}
    H_s &= \sum_k \begin{pmatrix}
        a_k\\b_k
    \end{pmatrix}^\dagger \left[\cH^s_x\sigma^x + \cH^s_y\sigma^y+\cH^s_z\sigma^z\right]\begin{pmatrix}
        a_k\\b_k
    \end{pmatrix}
\end{align}
where $\cH^s_z = m$, and the other coefficients are straightforward to compute but will not be needed here. Unlike the two-band Chern insulator model, $\cH^s_z$ is independent of the momenta. Hence, there exists a globally well-defined Chern connection, leading to zero ground-state Chern number for any $m\neq 0$. When $m=0$, the theory is gapless. Stated differently, time reversal symmetry acts as $\psi(r)\rightarrow(-)^{i+j}\psi(r)^\dagger$, which leaves the Hamiltonian \eqref{StagH} invariant. This invariance guarantees the vanishing of the Chern number in the ground state.

These observations can be rephrased in terms of fermionic doublers. The continuum theory of two Dirac fermions coupled to a $\U(1)$ gauge field is time-reversal invariant, provided that the two masses are equal and opposite. On the lattice Hamiltonian side, the staggered fermion formulation reduces the doublers from four to two but does not eliminate them entirely. Due to the staggered mass, these two modes have equal and opposite mass, making the theory time-reversal invariant. To construct a time-reversal-breaking Hamiltonian theory of a single doubler-free fermion coupled to a $\U(1)$ gauge field, we use Wilson fermions. The Wilson term removes all doublers in the Hamiltonian formulation and breaks time-reversal symmetry explicitly. This is consistent with well-known results from the Lagrangian formulation on the topological phases of Wilson fermions, as we will establish shortly.

\section{Topological Phases of $N_f=1$ QED$_3$}\label{sec:Nf=1}
In this section, we analyze the topological phase diagram of QED$_3$ with $N_f=1$ Wilson fermion. The Hamiltonian $H_{\rm QED_3}= H_f+H_m+e^2H_E-\frac{1}{e^2}H_B$ for this theory in temporal gauge is given below
\begin{align}
    H_B & =  \sum_r\cos \sfB(r)\no\\
    H_E & = \frac{1}{2}\sum_{k,\;r}\sfE_k(r)^2\no\\
    H_m & = (m+2R)\sum_r\psi_r^\dagger\gamma^0\psi_r\no\\
    H_f &= \frac{1}{2} \sum_{r} \left[\psi^\dagger_{r}\gamma^0\left(i\gamma^k + R \right) \psi_{r+\hat{k}}\sfU_{k}(r) + \textit{h.c.}\right]\no 
\end{align}
where the link variables are $\sfU_{x}(r) = e^{iX_r}$, $\sfU_{y}(r) = e^{iY_r}$ and $\sfB(r)=X_{r}+Y_{r+\hat{x}} - X_{r+\hat{y}} - Y_{r}$. Moreover, we have the canonically conjugate electric and magnetic fields
\begin{align}
    [\sfE_k(r), \sfU_\ell(r')] = e\delta_{k,\ell}\delta_{r,r'}\sfU_k(r)
\end{align}
For future convenience, let us define the lattice derivative operators:
\begin{align}
    \Delta_{k, r}^+ f(r) &= f(r+\hat{k})-f(r) &\Delta_{k, r}^- f(r) &= f(r)-f(r-\hat{k})
\end{align}
The Gauss-law constraint can be phrased in terms of a local operator $\sfG(r)$ that commutes with the full Hamiltonian $H_{\rm QED_3}$:
\begin{align}
    \sfG(r) = \Delta^{-}_{k,r}\sfE_k(r) - \cQ(r) 
\end{align}
where $\cQ(r) = \psi_r^\dagger\psi_r$ is the local charge-density operator. The simultaneous diagonalizability of $H$ and $\sfG$ implies that the Hilbert-space decomposes into eigenspaces of the Gauss-law operator $\sfG$. Our goal will be to project onto the zero-eigenspace in this decomposition. In other words, 
\begin{align}
    \ket{\psi}\in\cH_\text{phys}\iff \sfG(r)\ket{\psi} = 0\text{ for all }r.
\end{align}
Averaging over the group of all Gauss' law operators, we define a projector $\amsmathbb{P}$ onto the physical Hilbert space $\cH_\text{phys}$, which commutes with the full Hamiltonian. A state $\ket{\psi}\in\cH_\text{phys}$ iff $\amsmathbb{P}\ket{\psi} = \ket{\psi}$. 

Since we are interested in quantum simulations, we will often be interested in $\integers_N$ truncations of the above gauge theory. Explicitly, we use an exponentiated form of the Gauss constraint, defined in terms of $\cG_r = e^{i\frac{2\pi}{N}\sfG(r)}$, of which $L^2 -1$ are independent due to the global constraint. The group $\amsmathbb{G}^{(N)}$ (resp. $\amsmathbb{G}$) of Gauss' law operators is $\integers_N^{L^2-1}$ (resp. $\U(1)^{L^2-1}$). Hence, we may define the normalized projector $\amsmathbb{P}$ onto the gauge-invariant subspace $\cH_\text{phys}$ as
\begin{align}\label{projZ2}
    \amsmathbb{P} = \frac{1}{\sqrt{\abs{\amsmathbb{G}^{(N)}}}}\sum_{\cG\in\amsmathbb{G}^{(N)}}\cG.
\end{align}

Our strategy for the remainder of this section is as follows. First, we explain why the gauge background is trivial (i.e. $\sfU_k = 1$ for $k=x, y$) at weak-coupling, and solve the free-fermion theory, which exhibits non-trivial topological phases. Second, we uplift this solution to be gauge-invariant using the Gauss-projector $\amsmathbb{P}$, which does not change the spectrum or location of gapless transition points. Next, we explain why this solution represents the true ground state in the so-called \textit{trivial-flux sector}, to be defined below. In addition, we prove that the topological phase diagram that we find here is consistent both with the lattice Lagrangian formulation and the continuum limit. We close the section with an analysis of time-reversal symmetry.
\subsection{Flux-symmetry at weak-coupling}
\label{sec:FluxSym}
To analyze the spectrum and topological phases at weak coupling, we identify the maximally commuting part of the Hamiltonian, which we regard as the ``unperturbed" Hamiltonian $H$ with perturbation $H'$:
\begin{align}
    H &= H_B + H_f+H_m, &H' &= H_E.
\end{align}
Since the terms in $H$ are mutually commuting, we may simultaneously diagonalize the above Hamiltonian. Furthermore, the weakly-coupled limit has an emergent $\U(1)^x\times \U(1)^y$ global symmetry that greatly assists in the analysis of topological phases. In particular, the non-contractible Wilson lines, given by:
\begin{align}
    \cW_x &= \prod_{x} \sfU_x(x, 0) &\cW_y &= \prod_y \sfU_y(0, y)
\end{align}
commute with the unperturbed Hamiltonian $H$, i.e. $[\cW_k, H] = 0$. 

In the discussions to follow, we focus on the \textit{trivial-flux sector} defined by $(\cW_x, \cW_y) = (1,1)$. The ground-state in this superselection sector exhibits a non-trivial topological phase diagram, which we show is consistent with the IR phase diagram in the lattice Lagrangian formulation as well as the continuum limit. Therefore, in the thermodynamic limit, the ground state of the trivial-flux sector is expected to converge to the unique ground state of the continuum theory (see Sec.~\ref{sec:TruncationFiniteSize}). 

To show that the flux constraint is compatible with gauge symmetry, we show $[\cW_k, \sfG(r)] = 0$ for any $r$ assuming periodic boundary conditions.
\begin{align}
    &[\sfG(x', 0), \cW_x] \no\\
    &= \sum_{x}\left[\sfE_x(x,0) - \sfE_x(x-1, 0), \prod_{x'}\sfU_x(x', 0)\right] \no\\
    &= \sum_x[\sfE_x(x,0), \sfU_x(x,0)]\prod_{x'\neq x}\sfU_x(x',0) &\no\\&- \sum_x[\sfE_x(x-1,0), \sfU_x(x-1,0)]\prod_{x''\neq x-1}\sfU_x(x'',0) = 0,
\end{align}
which vanishes due to periodic boundary conditions. This argument holds for any lattice size $L$ and both $\integers_N$ or $\U(1)$ gauge groups provided the gauge fields satisfy periodic boundary conditions. A parallel argument works for $\cW_y$.

\subsection{The Hamiltonian in the trivial flux sector}\label{sec:TrivialFlux}
We take the spatial lattice to be $L\times L$ (with periodic boundary conditions), and each site contains a single Wilson fermion, which corresponds to two complex degrees of freedom. In total, we have $2L^2$ complex degrees of freedom. Explicitly, take $\Psi = (\psi_{1,1}, \psi_{2,1}, \dots, \psi_{L,1}, \dots, \psi_{1, L}, \psi_{2,L},\dots, \psi_{L, L})^T$, which is a $2L^2$-vector. In this notation, the Hamiltonian may be put in the following form:
\begin{align}\label{HR}
    H_f = \Psi^\dagger \cH_R\Psi
\end{align}
Hence, the Hamiltonian $H_f$ is associated to a Hermitian matrix in real-space $\cH_R$ of size $2L^2\times 2L^2$. Since the above Hamiltonian, is gauge-invariant, i.e. $[\amsmathbb{P}, H_f]=0$, the $2L^2$ real eigenvalues must also be gauge-invariant. In other words, when this Hamiltonian is diagonalized with a fixed gauge field configuration 
\begin{align}
    (X, Y)\coloneqq\{(X_{x,y}, Y_{x, y})|i, j = 0, \dots, (L-1)\},
\end{align}
we find $2L^2$ real eigenvalues $E^{\pm}_{x, y}$ which are functions of the gauge-equivalence classes or orbits $[(X, Y)]$, defined by the equivalence relation on the gauge Hilbert space $\cH_G$
\begin{align}\label{equivalenceRel}
    (X, Y)\sim (X', Y')\iff \exists \;\cG\in \amsmathbb{G}:\;\ket{X, Y} = \cG\ket{X', Y'}.
\end{align}
Written explicitly, we have:
\begin{align}
    E^\pm_{x, y}([(X, Y)]) = E^{\pm}_{x, y} (X, Y)
\end{align}
Hence, to solve for the ground-state in the trivial-flux sector with $\cW_x = \prod_x e^{iX_{x, 0}} = 1$ and $\cW_y = \prod_ye^{iY_{0,y}} = 1$, we simply need to identify \textit{one representative} $(X^*, Y^*)$ of the gauge-orbit $[(X^*, Y^*)]$ which minimizes the fermionic energy at half-filling. Once such an element is identified, we can easily reconstruct the gauge-invariant solution $\amsmathbb{P}[\ket{f(X^*, Y^*)}\otimes \ket{X^*, Y^*}]$. 

In the following sub-subsection, we present an argument to show that this representative may be chosen to be $(X^*, Y^*) = (0^L, 0^L)$ provided we restrict to the superselection sector of $H$ labeled by $(\cW_x, \cW_y)=(1, 1)$. We emphasize that this is a valid choice at \textit{weak-coupling} because $[\cW_x, H] = [\cW_y, H] = 0$, where $H = H_f + H_m - \frac{1}{e^2}H_B$. For larger couplings, this approximate symmetry does not exist because $[H_E, \cW_k]\ne 0$ is no longer small. This argument is exemplified in the simplest possible setting of a $2\times 2$ spatial lattice with $\integers_2$ gauge fields. We have verified the correctness of this choice for larger systems $L=4, 6$ as well as higher truncations $N
=3, 4$. In the final sub-subsection, we use this gauge choice to rewrite the Hamiltonian in momentum-space, which allows us to compute the fermionic ground state and hence characterize topological phases.

\subsubsection{Details on the $2\times 2$ case with $\integers_2$ gauge fields}\label{OrbitGS}
The gauge-parts of physical states are labeled as $\ket{X_{11}, X_{21}, X_{12}, X_{22}; Y_{11}, Y_{21}, Y_{12}, Y_{22}}$, where $e^{i\pi X_{ij}}, e^{i \pi Y_{ij}}\in\integers_2 \cong \{-1, +1\}$\footnote{Note that we have moved to a convention where $X_{x,y}, Y_{x, y}$ are $0$ or $1$ since the gauge group is $\integers_2$.}, which may be regarded as eigenvalues in the Pauli-$Z$ basis. The Gauss constraint may be phrased in terms of the exponentials $\cG_r$ of the operators $\sfG(r)$, represented by Pauli-$X$ operators (the superscript denotes the orientation of the links):
\begin{align}
    \cG_{ij} = (\sigma^x)_{ij}^{(x)}(\sigma^x)_{i-1,j}^{(x)}(\sigma^x)_{ij}^{(y)}(\sigma^x)_{i,j-1}^{(y)}e^{-i\pi\psi_{i,j}^\dagger\psi_{i,j}}
\end{align}
The group of $\integers_2$-Gauss' law operators $\amsmathbb{G}^{(2)}$ is generated by $(L^2 -1)$ of the Gauss' law operators, since the last vertex imposes a redundant constraint. Hence, we have
\begin{align}\label{G2}
    &\amsmathbb{G}^{(2)} = \expval{\cG_{11}, \cG_{12}, \cG_{21}} \no\\&= \{\mathbb{1},\cG_{11}, \cG_{12}, \cG_{21}, \cG_{11}\cG_{12}, \cG_{12}\cG_{21}, \cG_{11}\cG_{21}, \cG_{11}\cG_{12}\cG_{21}\} \no\\&\cong \integers_2\times\integers_2\times\integers_2
\end{align}
This will be crucial for diagonalizing the real-space Hamiltonian $\cH_R$ in \eqref{HR}.

As previously mentioned, we can only determine a set of gauge configurations that minimize the fermionic energy up to the equivalence relation \eqref{equivalenceRel}. This may be explicitly confirmed in the $2\times2$ $\integers_2$ case by looping over all allowed gauge configurations, which yields a set of $8$ gauge configurations that {are} related precisely by elements of $\amsmathbb{G}^{(2)}$ --- all of which minimize the energy of the fermionic ground state. Explicitly, label the sites by pairs $r=(x, y)$ with $x,y =1,2$. It is convenient to define
\begin{align}
    M_x & = \gamma^0\left(i\gamma^x + R\,\mathbb{1}_2 \right),& M_y & = \gamma^0\left(i\gamma^y + R\,\mathbb{1}_2  \right).
\end{align}
Furthermore, we define
\begin{align}
    \sfU_{x}(x,y)&=e^{iX_{x,y}} &\sfU_{y}(x,y)&=e^{iY_{x, y}},
\end{align}
and $M_x^{x,y} = M_x e^{iX_{x, y}}, M_y^{x, y}= M_y e^{iY_{x, y}}$. The hopping Hamiltonian, including contributions from the Wilson term, $\cH_{\text{hop}}$ is:
\begin{align}
     \frac{1}{2}\begin{psmallmatrix}
        0 & M_x^{11}+(M_x^{21})^\dagger&M_y^{11}+(M_y^{12})^\dagger & 0\\
         M_x^{21}+(M_x^{11})^\dagger & 0& 0& M_y^{21}+(M_y^{22})^\dagger\\
          M_y^{12}+(M_y^{11})^\dagger& 0 &0 &M_x^{12}+(M_x^{22})^\dagger \\
          0& M_y^{22}+(M_y^{21})^\dagger &M_x^{22}+(M_x^{12})^\dagger  &0
    \end{psmallmatrix}.
\end{align}
where $\Psi = (\psi_{11}, \psi_{21}, \psi_{12}, \psi_{22})$, $\cH_R = \cH_\text{hop} + \cH_\text{mass}$ and 
\begin{align}
    \cH_{\text{mass}} & = (m+2R) \Psi^\dagger \left[\mathbb{1}_4\otimes \sigma^3\right] \Psi
\end{align}
Since we are interested in the weak-coupling limit, where $e^2\rightarrow 0$, it is crucial to include the kinetic term for the magnetic field. 
\begin{align}
    -\frac{1}{e^2}\cH_\text{mag} =-\frac{1}{e^2}\Big[&\cos\bigl(\pi (X_{21} + Y_{21} - X_{12} - Y_{12})\bigr)\no\\
    &+ \cos\bigl(\pi(X_{21} + Y_{12} - X_{22} - Y_{21})\bigr)\no\\
&+ \cos\bigl(\pi(X_{12} + Y_{22} - X_{21} - Y_{12})\bigr)
\no\\&+ \cos\bigl(\pi(X_{22} + Y_{12} - X_{21} - Y_{22})\bigr)\Big]\mathbb{1}_8
\end{align}
The role of this term is to penalize those plaquettes that do not satisfy $\cos\sfB = 1$, since they are suppressed at weak coupling. For a $\integers_N$ gauge theory, the electric term may be dropped since it is $\cO(e^2)$, whilst the magnetic term above is $\cO(e^{-2})$ and the fermionic terms are $\cO(1)$.\footnote{This heuristic holds only when the theory is gapped. } Stated differently, we are restricting to the unperturbed problem $H = H_f+H_m - \frac{1}{e^2}H_B$.

Lastly, we project to the trivial flux sector where $\cW_x = \cW_y = 1$. This is valid as we are in the weak coupling sector, where these operators commute with the Hamiltonian $H$. Diagonalizing $\cH_R$ this Hamiltonian by looping over all $2^8$ gauge-configurations, we find that the following $8$ gauge-states are ``degenerate" (in the sense that gauge-redundant due to the equivalence relation \eqref{equivalenceRel}) and minimize the fermionic energy for any value of $(m+2R)$:
\begin{align}
    \bigl\{
\ket{00000000},\;
\ket{00001111},\;
\ket{00110101},\;
\ket{00111010},\;\no\\
\ket{11000101},\;
\ket{11001010},\;
\ket{11110000},\;
\ket{11111111}
\bigr\}
\end{align}
It is straightforward to verify these $8$ elements belong to a single gauge-orbit and are related by the $8$ elements of $\amsmathbb{G}^{(2)} = (\integers_2)^3$ in \eqref{G2}. Hence, in the ground state of the trivial flux sector, one may set the gauge fields to zero \textit{up to gauge transformations}. With the above understanding, we may pick a gauge where $X_{r} = Y_{r} = 0$ since the spectrum itself is invariant under gauge transformations of this configuration. 

We have verified this finding for $L=4,6$ and $N=3, 4$ with a similar approach, but do not spell out these examples here for simplicity. We emphasize here that this gauge choice does \textit{not} trivialize gauge-dynamics, as we eventually also incorporate the effects of the perturbation $H_E$, which does \textit{not} commute with $H$ or $\cW_k$. However, at this stage, in the weak coupling limit, the gauge fields and fermions decouple up to gauge transformations, which allows for an exact solution.

\subsubsection{The $L\times L$ case in momentum space}\label{App:Hamiltonian}
Having fixed the gauge-background $(X^*, Y^*) = (0^L, 0^L)$ in the trivial flux sector, we now use the Fourier transform of the fermions to rewrite the Hamiltonian, which will allow us to solve for the ground-state in the trivial-flux sector at weak-coupling. Explicitly:
\begin{align}
    \psi_{r}& = \frac{1}{\sqrt{N}} \sum_{k\in\cB}e^{ik\vdot r}\psi_k,  
\end{align}
where $\cB$ is the Brillouin zone. This implies
\begin{align}
    H_f &= \half \sum_{r}  \Bigg[\psi_{r+\hat{x}}^\dagger M_x\psi_{r} + \psi_{r+\hat{y}}^\dagger M_y\psi_{r} \Bigg] + h.c.\no\\&= \half\sum_k \psi_k^\dagger \Bigg[e^{-ik_x} M_x+e^{-i k_y} M_y +h.c.\Bigg]\psi_k
\end{align}
Decomposing $M_x$ and $M_y$ into the Pauli matrices, we get three distinct contributions
\begin{align}
    \cH_x&=\sin k_x&\cH_y&=\sin k_y&\cH_z^{(1)}&=R\cos k_x+R\cos k_y
\end{align}
The mass term takes the following form
\begin{align}
    H_m &= (m+2R)\sum_r \psi_{r}^\dagger\sigma^z\psi_{r} = (m+2R) \sum_{k} \psi_k^\dagger \sigma^z \psi_k
\end{align}
which gives a second contribution to the $z$-component
\begin{align}
    \cH_{z}^{(2)} &= (m+2R)\sigma^z
\end{align}
where we now define $\cH_z = \cH_{z}^{(1)}+\cH_{z}^{(2)}$. To summarize the Hamiltonian in momentum space is 
\begin{align}\label{HNf=1}
    H &= \sum_{k}\psi_k^\dagger\cH(k_x, k_y)\psi_k,
\end{align}
with $\cH(k_x, k_y) = \cH_x\sigma^x+\cH_y\sigma^y+\cH_z\sigma^z$. From here on, let us define $M\equiv m+2$ and set $R=1$. Then we have:
\begin{align}\label{shiftMass}
    \cH_z = M +\cos k_x + \cos k_y.
\end{align}
This derivation and the argument for setting the gauge fields to zero up to gauge transformations fill the missing link between the low energy limit of Wilson fermions coupled to a $\U(1)$ gauge field and the Chern insulator Hamiltonian suggested in \cite{Sen:2020srn}. In fact, $\cH$ is simply the QWZ model \cite{Qi:2006xub} in condensed matter. 

\subsection{The unperturbed trivial-flux spectrum}
The Hamiltonian may be diagonalized by a unitary transformation:
\begin{align}
    \cH(k) = \cU(k)^\dagger \cH^{(D)}(k) \cU(k) 
\end{align}
where $\cH^{(D)} = \text{diag}\left(E_-(k), E_+(k)\right)$. It is convenient to define
\begin{align}\label{psiPM}
\Tilde{\psi}_k = \cU(k) \psi_k = \begin{pmatrix}
    \psi_{k-}\\\psi_{k+}
\end{pmatrix}
\end{align}
Then the Hamiltonian takes the simple form
\begin{align}
    H = \sum_k \left[E_-(k)\psi_{k-}^\dagger\psi_{k-}+E_+(k)\psi_{k+}^\dagger\psi_{k+}\right]
\end{align}
where $E_{\pm}(k_x,k_y) = \pm \sqrt{\cH_x^2+\cH_y^2+\cH_z^2} = \pm\abs{\cH}$, viewing $\cH=(\cH_x,\cH_y,\cH_z)$ as a vector. The band gap between the two levels is 
\begin{align}
    \Delta(k_x,k_y) &= E_+(k_x,k_y) - E_-(k_x,k_y) = 2\abs{\cH}.
\end{align}
The eigenstate for the lower band is 
\begin{align}
    \ket{u_-^{(1)}}& = \frac{1}{\cN_{1-}}\begin{pmatrix}
        \cH_z - \abs{\cH}\\
        \cH_x+i\cH_y
    \end{pmatrix}
\end{align}
which is singular when $\cH_x = \cH_y = 0$ and $\cH_z>0$ ($\cN_{1-}$ is a normalization constant). This occurs when 
\begin{align}
    (k_{px},k_{qy})&= \left(p\pi,q\pi\right)\text{ with }p,q\in\{0,1\}
\end{align}
and 
\begin{align}
    \cH_z(k_{px}, k_{qy}) &= M+(-1)^p+(-1)^q > 0
\end{align}
At this singularity in the Brillouin zone $\cB$, we must use an alternative eigenstate that is non-singular and gauge-equivalent to $u_-^{(1)}$, i.e.
\begin{align}
    \ket{u_-^{(2)}}& = \frac{1}{\cN_{2-}}\begin{pmatrix}
        \cH_x+i\cH_y\\
        \cH_z + \abs{\cH}
    \end{pmatrix} = e^{i\lambda_-(k_x,k_y)}\ket{u_-^{(1)}}
\end{align}
where  $\cN_2$ is a normalization constant and
\begin{align}
    e^{i\lambda_-(k_x,k_y)} &= \left({\frac{\cH_z+\abs{\cH}}{\cH_x+i\cH_y}}\right){\abs{\frac{\cH_z+\abs{\cH}}{\cH_x+i\cH_y}}}^{-1}
\end{align}
This wavefunction is complementary in the sense that it is singular when $\cH_x=\cH_y=0$ and $\cH_z<0$. An analogous expression exists for $\lambda_+$, but we will not need its explicit form.

The fermionic vacuum-state $\ket{f}$ is represented as
\begin{align}\label{FockVac}
    \ket{f} &  = \bigotimes_{k\in\cB} \psi_{k-}^\dagger\ket{\emptyset} = \bigotimes_{n,m = 0}^{L-1}\ket{u_-(k_{n,m})}
\end{align}
with $\ket{\emptyset}$ being the state with no particles and $ k_{n,m} = \left({2\pi n}/{L}, {2\pi m}/{L}\right)$ where $n,m = 0, 1,\dots, (L-1)$. In equation \eqref{FockVac}, the subtlety regarding whether we use $u_-^{(1)}$ or $u_-^{(2)}$ is left implicit.

\subsection{Calculation of the Chern number}\label{WarmUp}
To characterize the IR topological phases, we use the first Chern number of the following $\U(1)$ connection and curvature (not to be confused with the gauge field)\footnote{Note the independence of $b$ on the index $\ell$, which follows from $\U(1)$ gauge-invariance} 
\begin{align}
    a_{i}^{(\ell)} &= -i\bra{u_-^{(\ell)}}\partial_{k_i}\ket{u_-^{(\ell)}}, & b_i = (\nabla\times{a^{(\ell)}})_i.
\end{align}
A few comments are in order. 
\begin{enumerate}
    \item When the mass $M>2$, $\inf_{\cB} \cH_z>0$ for any choice of the momenta and the $\ket{u_-^{(2)}}$ is globally well-defined: this phase is topologically trivial.
    \item Similarly, when $M<-2$, $\sup_{\cB} \cH_z < 0$ for any choice of momenta and $\ket{u_-^{(1)}}$ is globally well-defined: this phase is topologically trivial.
    \item On the other hand, when $\abs{M}<2$, we expect to see topological phases with non-zero Chern numbers since the two eigenstates $\ket{u_{-}^{(1,2)}}$ are patched together by the non-trivial gauge transformation $\lambda$. 
\end{enumerate}

To identify the topological phases, we note the following sign-structure for $\cH_z$ at the singular points $(k_{px}, k_{qy})$.
    \begin{enumerate}
    \item $\cH_z(k_{0x}, k_{0y}) = M+2 >0$ for any $M>-2$.
    \item $\cH_{z}(k_{1x}, k_{0y}) =\cH_{z}(k_{0x}, k_{1y})= M>0$ if $M > 0$. 
    \item $\cH_{z}(k_{1x}, k_{1y}) = M - 2>0$ for any $M > 2$.
    \end{enumerate}

Using the above inputs, we may evaluate the Chern number. For any $-2<M<0$, we must use the eigenstate $u_-^{(2)}$ in a neighborhood of the $(p,q)=(0,0)$ point, and $u^{(1)}_-$ in its complement in $\cB$. Calling the common boundary contour $\gamma$, the first Chern number of $b$ is
    \begin{align}\label{ChNo}
        c_1[b] & = \int_\cB \frac{b}{2\pi}= \frac{1}{2\pi}\int_\gamma\left(-a^{(1)} + a^{(2)}\right) =\int_\gamma \frac{d\lambda}{2\pi}
    \end{align}
    where we have used Stokes' theorem, and the fact that the two connections differ by the gauge-transformation $\lambda$. Parametrizing the curve $\gamma$ by an angle $\theta\in[0,2\pi]$,
    \begin{align}
        c_1[b] = \frac{\lambda(2\pi) - \lambda(0)}{2\pi}.
    \end{align}
    Supposing that $\gamma$ is a small loop around $(k_x,k_y) = \left(0, 0\right)$, we Taylor expand as follows
    \begin{align}
        \cH_x& =  k_x + \cO(k_x^2), &\cH_y& =  k_y + \cO(k_y^2)
    \end{align}
    Since \eqref{ChNo} is manifestly coordinate invariant, we are free to evaluate it in any choice of coordinates. Hence:
    \begin{align}
        e^{i\lambda} = \frac{\abs{\cH_x+i\cH_y}}{\cH_x+i\cH_y} = \frac{\abs{k_x + i k_y}}{k_x + i k_y} + \cO(k^2) = e^{-i\theta} + \cO(k^2)
    \end{align}
    where we have switched to polar coordinates $(k,\theta)$. This yields $\lambda = -\theta$. Hence,
    \begin{align}
        c_1[b] = -1 \text{ for any }M\in(-2,0).
    \end{align}
    For any $0<M<2$, we must use the eigenstate $u_-^{(1)}$ in a neighborhood $\cB^{(1)}$ of the $(p,q)=(1,1)$ point in $\cB$, and $u^{(2)}_-$ in the complement. The above argument applies except that the orientation of $\gamma$ since we take it to be a circle around the $(p,q) = (1,1)$ point rather than the $(0,0)$ point. Hence,
        \begin{align}
            c_1[b] = +1 \text{ for any }M\in(0,2).
        \end{align}

In summary, the first Chern number is 
\begin{align}
    c_1[b] & = \int_\cB \frac{b}{2\pi}=\begin{cases}
        -1,&\text{ if }M\in(-2,0)\\
        +1,&\text{ if }M\in(0,+2)\\
        0,&\text{ otherwise }
    \end{cases}
\end{align}
In terms of the many-body fermionic solution $\ket{f_g}$ subject to the gauge background $\ket{g}=\bigotimes_{k, r}\ket{\sfU_k(r)=1}$, the gauge-invariant solution is $\amsmathbb{P}\ket{f_g, g}$, which has the same energy as $\ket{f_g,g}$ since $[H,\amsmathbb{P}] = 0$. Real space diagonalization on small lattices shows that $\sfU_k = 1$ (see Sec.~\ref{OrbitGS}) in the vacuum up to gauge-equivalent configurations related by Gauss' law \textit{provided}: $(\cW_x, \cW_y)=(1,1)$ \textit{and} we include the plaquette term $-\frac{1}{e^2} \sum_r\cos \sfB(r)$ at weak-coupling. In fact, the $(1,1)$ sector contains the true weak-coupling vacuum in the thermodynamic limit.
\begin{figure}[b]
    \centering
    \includegraphics[width=0.72\linewidth]{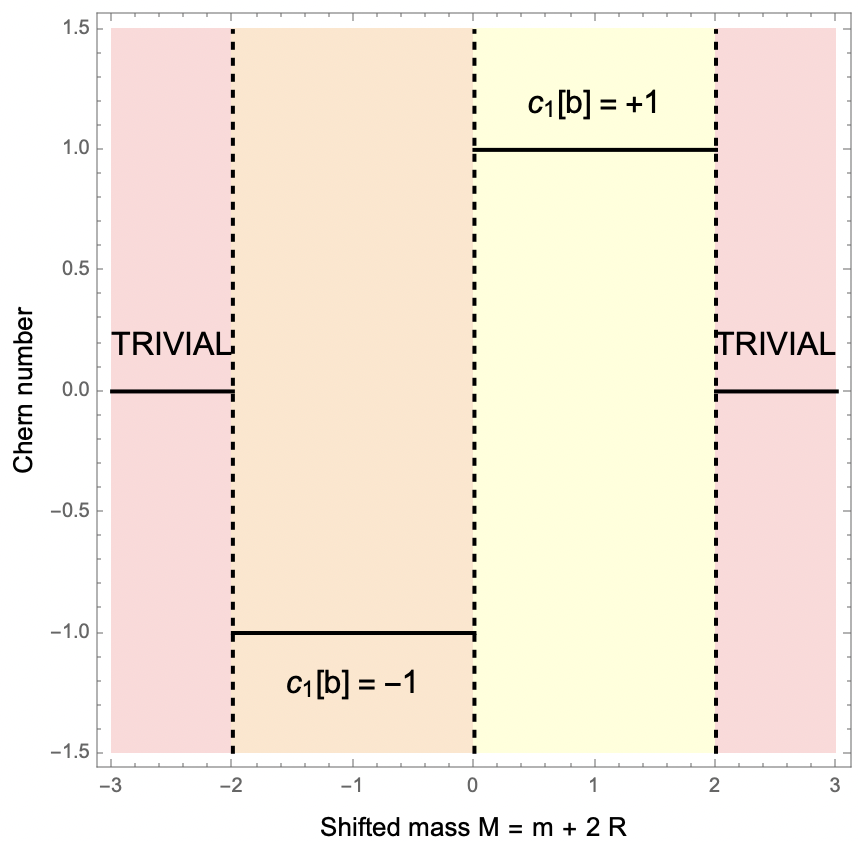}
    \caption{ Chern number (solid black line) vs. the shifted mass $M$ for $R=1$. The pink regions are trivial insulator phases, the orange one has $c_1[b] = -1$, while the yellow one has $c_1[b] = +1$.\label{fig:ChNo}}
    
\end{figure}

\subsection{Consistency with the lattice Lagrangian formulation}\label{App:ConsistencyWithLagrangian}
In this subsection, we establish a relation between the vacuum Chern number that we have computed in the Hamiltonian formulation (i.e. with time being continuous but space being discrete) with the Chern-Simons level obtained in the Lagrangian formulation (i.e. with both time and space discrete) \cite{Golterman:1992ub, Sen:2020srn}. We find full agreement with Fig.~\ref{fig:ChNo}, including the locations of the transitions at $M=0, \pm 2$. Starting with the Chern-Simons level derived in the Lagrangian formulation with spatial lattice spacing $a$ and time lattice-spacing $a_0$, we consider the limit $a_0\rightarrow 0$, derived in equation 2.9 of \cite{Sen:2020srn}. In our conventions with $a=1$, this equation reads
\begin{align}
    c(m,a_0)&=\frac{1}{2}\vast[(-)^0\frac{m}{\abs{m}}+ (-)^1 \left[2\frac{m+2}{\abs{m+2}} + \frac{m+\frac{2}{a_0^2}}{\abs{m+\frac{2}{a_0^2}}}\right]\no\\& + (-)^2\left[\frac{m+4}{\abs{m+4}} + 2 \frac{m+2+\frac{2}{a_0^2}}{\abs{m+2+\frac{2}{a_0^2}}}\right]\no\\&+ (-)^3\frac{m+2+\frac{4}{a_0^2}}{\abs{m+2+\frac{4}{a_0^2}}}\vast]
\end{align}
Using $M = m + 2$, we have
\begin{align}
    \lim_{a_0\rightarrow 0}c(M, a, a_0) &= \frac{1}{2}\left[\frac{M-2}{\abs{M-2}} -2\frac{M}{\abs{M}}+\frac{M+2}{\abs{M+2}}\right] \no\\ &= \begin{cases}
        0,&M<-2\\
        +1,&M\in(-2,0)\\
        -1,&M\in(0,2)\\
        0,&M>2
    \end{cases}
\end{align}
which reproduces $c_1[b]$ (up to an arbitrary sign that is fixed by the orientation of the integration contour). This can be summarized as
\begin{align}
    c_1[b]& = - \lim_{a_0\rightarrow 0}c(m, a, a_0).
\end{align}
This confirms that the topological phases that one obtains in the deep IR from the Lagrangian formulation agree with what we find using the Hamiltonian. Importantly, the topological transition that we have at $m=6$ for $a_0\neq 0$ in the Lagrangian formulation disappears. Satisfactorily, we can match the Lagrangian answer with the full continuum limit. For this purpose, we restore the spatial lattice-spacing $a$. 
\begin{align}
    c(m, a, 0) &= \frac{1}{2}\left[\frac{m}{\abs{m}} -2\frac{m+\frac{2}{a^2}}{\abs{m+\frac{2}{a^2}}}+\frac{m+\frac{4}{a^2}}{\abs{m+\frac{4}{a^2}}}\right] \no\\
    &\xrightarrow{a\rightarrow 0} \frac{1}{2}\left(\frac{m}{\abs{m}} - 1\right)  = \begin{cases}
        0,& m>0\\
        -1,&m<0
    \end{cases}
\end{align}
This is precisely the Chern-Simons level for Pauli-Villars regulated $(2+1)d\;\U(1)$ gauge theory with $N_f = 1$ Dirac fermion and negative Pauli-Villars mass.

\subsection{Time-reversal symmetry}
In this section, we analyze the time-reversal transformations of a single Wilson fermion coupled to a $\U(1)$ gauge field. Since the gauge kinetic terms are invariant, we restrict our attention to the fermionic terms. We take the anti-unitary time-reversal operator to be 
\begin{align}
    \cT = i\sigma^y \cK
\end{align}
where $\cK$ is complex-conjugation operator. Under $\cT$-transformations, $(k_x, k_y)\rightarrow(-k_x, -k_y)$,  $\gamma^\mu\rightarrow-\gamma^\mu$. Since we work in the temporal gauge and the spatial components of the gauge fields have the same transformation as the momenta, i.e. $(X, Y)\rightarrow(-X, -Y)$, and hence do not play a role in this discussion.

For the purposes of this discussion, we may write the Dirac and Wilson parts of the Hamiltonian in momentum space as: 
\begin{align}
    \cH_D &= \gamma^0(\gamma^i\sin k_i + m) = \sigma^i\sin k_i + m\sigma^z \no\\ \cH_W &= R(2+\cos k_x+ \cos k_y)\sigma^z
\end{align}
Then we have
\begin{align}
    \cT^\dagger \cH_D\cT &= \sigma^i\sin k_i - m\sigma^z, &\cT^\dagger \cH_W\cT = -\cH_W.
\end{align}
To summarize, $\cT$ effectively maps $(m, R)\rightarrow(-m, -R)$. Hence, the $N_f=1$ Wilson fermion theory explicitly breaks time-reversal symmetry, allowing for topological phases characterized by an integer Chern number.

\section{Phase Diagram of $N_f=2$ QED$_3$} \label{sec:Nf=2}
We consider the topological phase diagram of the two-flavor theory as a function of the masses, both at zero and at finite density. The gauge-kinetic contribution and arguments for fixing the gauge at weak coupling to $\sfU_k=1$ is identical to what we have previously described for the $N_f=1$ case (see Sec.~\ref{OrbitGS}). Hence, to analyze the phase diagram, we need only analyze the fermionic terms. At zero chemical potential, the fermionic Hamiltonian with $N_f = 2$ flavors coupled to a $\U(1)$ gauge field and relative chemical potential $\mu$\footnote{For the $N_f=2$ phase diagram at zero density, see Fig.~\ref{fig:ZeroDensityNf=2}, \cite{SM}.} is
\begin{align}
    H_f = &\frac{1}{2} \sum_{r} \left[\psi_a(r)^\dagger\gamma^0\left(i\gamma^k + R \right) \psi_a(r+\hat{k})\sfU_{k}(r) + h.c. \right]\no\\
    & +\sum_{r} M_{a} \psi_a(r)^\dagger \gamma^0 \psi_a(r)+\mu(\psi_1^\dagger\psi_1 - \psi_2^\dagger\psi_2).
\end{align}
where $a = 1, \dots, N_f$.  In the second line above, we have implicitly included the dependence on the Wilson couplings inside the shifted mass-matrix $M_{a} = m_{a} + 2 R$ (we set $R=1$ here). We will focus on the following special cases: $M_1=M_2=M$ (only a singlet mass) and $M_1=-M_2=M$ (only a triplet mass). We focus on the theory at half-filling, defined by $\frac{1}{4L^2} \sum_r \psi_a(r)^\dagger\psi_a(r) = \half$. The average occupation fraction $f_a(\mu, M)$ at half-filling serves as an indicator for metal-insulator transitions with 
\begin{align}
    f_a(\mu, M) & = \frac{1}{2L^2}\sum_{k\in\cB}\expval{\psi_a(k)^\dagger\psi_a(k)}_{\mu, M} 
\end{align}
where $\sum_a f_a(\mu, M) = 1$. 
\begin{figure}[b]
        \centering
        \includegraphics[scale=.7,width=.75\linewidth]{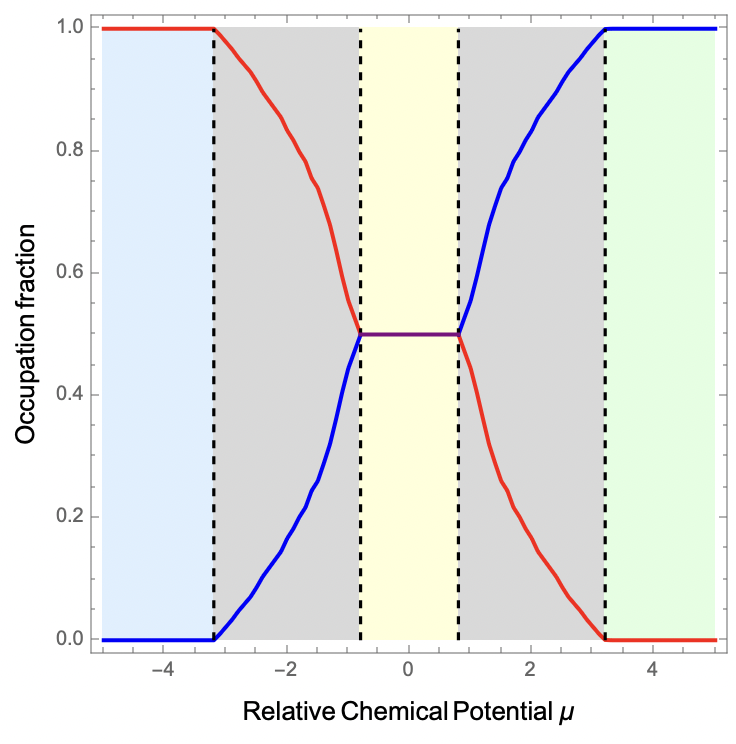}
        \caption{A plot of the average occupation fraction $f_a$ vs. $\mu$, where $a=1$ corresponds to $\uparrow$ (red) and $a=2$ corresponds to $\downarrow$ (blue). The purple segment in between corresponds to the regime where they red and blue coincide. Clearly, as we tune the chemical potential, the theory is driven to different phases. This plot was produced for the $N_f=2$ theory with equal masses $M=-1.2$ on a $32\times 32$ spatial lattice. 
        }
        \label{fig:OccNo}
\end{figure}
\subsection{Calculation of Chern Number}

\begin{figure*}[t]
\centering
\includegraphics[width=0.35\linewidth]{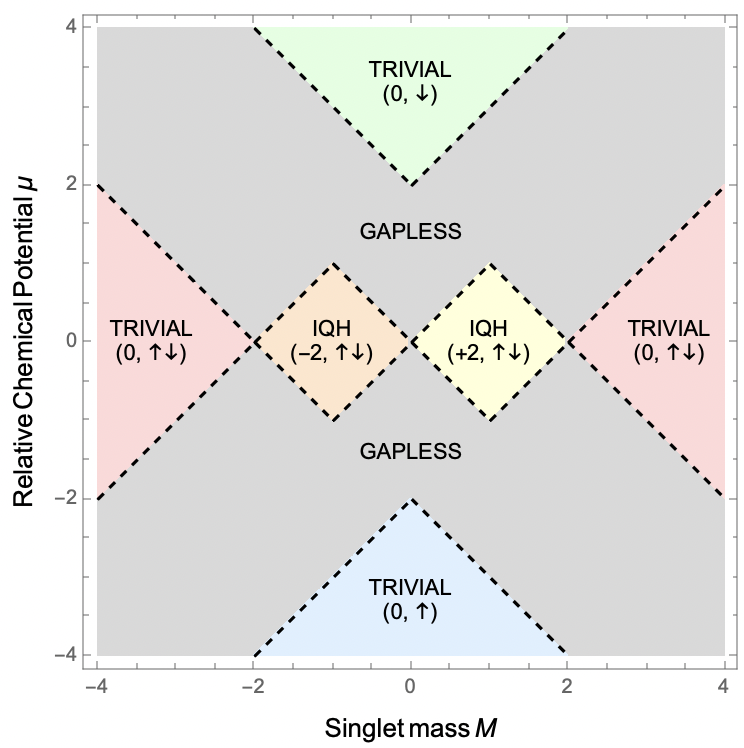}\hspace{5em}
\includegraphics[width = 0.35\linewidth]{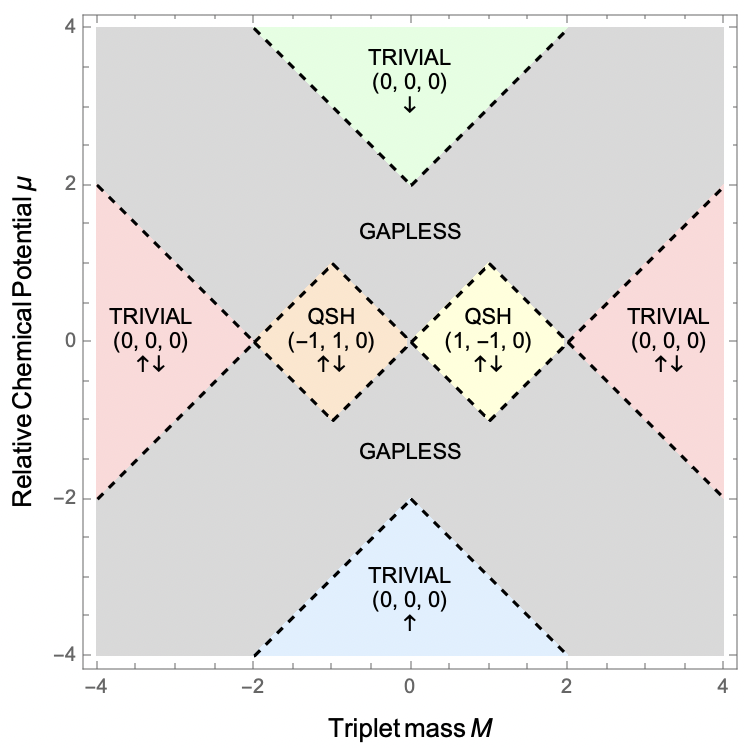}

\caption{The phase diagram of  $N_f=2$ Wilson fermions coupled to a $\U(1)$ gauge field for singlet (left) and triplet (right) mass configurations at weak coupling. In the left figure, the pairs are labeled as $(c_1[b], \expval{s})$. In the right figure, the triplets of Chern numbers are labeled as $(c_\uparrow, c_\downarrow, c_\text{tot})$. Here, ``$\uparrow\downarrow$" denotes zero average spin. The diagram highlights the existence of IQH phases for singlet masses and QSH phases for triplet masses. 
This plot was produced on a $16\times 16$ lattice using the analytical solution, which allows access to large system sizes.\label{fig:PhaseDiag}}
\end{figure*}

\subsubsection{Singlet mass}\label{App:ChernEqualM}
Since the structure of the unit cell is independent of the matter content, the Hamiltonian in Fourier space is: 
\begin{align}
    \cH(k)=\cH_x\Gamma^x+\cH_y\Gamma^y+\cH_z \Gamma^z+ \mu\; \sigma^z\otimes \mathbb{1}_2
\end{align}
where $\Gamma^k = \mathbb{1}_2\otimes \sigma^k$ and the functions $\cH_i(k)$ are as defined for the $N_f=1$. The eigenstates of $H_\text{Chem}$ are simply those of $\sigma^z$: $\{\ket{\uparrow}, \ket{\downarrow}\}$ corresponding to the energies $\pm \mu$. On the other hand, the eigenstates of $H$ are simply given as tensor-product states $\ket{s, u_\pm}$ with $s=\uparrow,\downarrow$. As before, we still have a gauge-ambiguity in the global definitions of $u_\pm(k)$.
The corresponding energies of the four bands are given by $E(\mu, k) = \pm \mu \pm \abs{\cH(k)}$, where the two choices of sign are independent. To map out the IR phase diagram, we need to know the structure of the vacuum at half-filling (see Fig.~\ref{fig:VacuumStructure}, Sec.~\ref{App:SupplementaryFigs}, \cite{SM}). In either regime with $\abs{\mu}\gg \sup\abs{\cH}$ or $\abs{\mu}\ll \inf\abs{\cH}$, the vacuum is gapped at half-filling. From the eigenvalues, we see
\begin{align}
    \ket{0} = \begin{cases}
        \bigotimes_{k\in \cB} \ket{\downarrow,u_-(k)}\otimes \ket{\downarrow,u_+(k)}, &\mu\gg \sup\abs{\cH}\\
        \bigotimes_{k\in \cB} \ket{\downarrow,u_-(k)}\otimes \ket{\uparrow,u_-(k)}, &|\mu|\ll \inf\abs{\cH}\\
        \bigotimes_{k\in \cB} \ket{\uparrow,u_+(k)}\otimes \ket{\uparrow,u_-(k)}, &\mu\ll -\sup\abs{\cH}
    \end{cases}
\end{align}
All of these regimes correspond to distinct gapped phases. This can be made precise by studying the vacuum expectation value of the average occupation fraction $f_a(\mu, M)$ as a function of $\mu$ for a fixed mass $M$. Fig.~\ref{fig:OccNo} shows three distinct insulating gapped phases (blue, yellow and green) separated by two metallic gapless (gray) crossover regions. The gapped phases may be topological or not: there exist integer quantum Hall (IQH) phases for the singlet case and quantum spin Hall (QSH) phases for the triplet case (see Fig.~\ref{fig:PhaseDiag}).
Repeating the analysis of Fig.~\ref{fig:OccNo} on various slices, we obtain the full phase diagram displayed in Fig.~\ref{fig:PhaseDiag}.  In the absence of a gauge field, the triplet mass case with $\mu = 0$ reduces to the BHZ model in condensed matter \cite{Bernevig:2006uvn}. 

We now evaluate the Chern number using the methods laid out in detail for the $N_f=1$ case in various parts of the full phase diagram, which was obtained by patching together a series of slices that look like Fig.~\ref{fig:OccNo}. This will give us a full understanding of the topological phase diagram in the $(M, \mu)$-plane, as summarized in Fig.~\ref{fig:PhaseDiag}. An analysis of the order of the metal-insulator transitions is considered in Sec.~\ref{Order} of the Suppementary Materials.

Let us first begin with the blue and green regions, of which it suffices to consider one. We focus on the green region in Fig.~\ref{fig:OccNo} exclusively, where the vacuum is
\begin{align}
    \ket{0}=\bigotimes_{k\in \cB} \ket{\downarrow,u_-(k)}\otimes \ket{\downarrow,u_+(k)}
\end{align}
We find that the Chern number of the vacuum in the green regime is always zero:
\begin{align}
     \frac{1}{L^2}C_\text{green} = c_1[b] - c_1[b] = 0
\end{align}
corresponding to $u_-$ and $u_+$ respectively (since the total Chern number of the two-band model is zero). However, if $\abs{M}<2$, each term is non-zero since $c_1[b]\neq 0$: this is a trivially gapped phase. The same logic applies to the blue region in Fig.~\ref{fig:OccNo}, with the only distinguishing feature being the sign of the flavor spins. 

In the intermediate gapped phase (which exists for $0<\abs{M}<2$: the yellow/orange regime in Fig.~\ref{fig:PhaseDiag}), the vacuum is given by 
\begin{align}
    \ket{0} &= \bigotimes_{k\in \cB} \ket{\downarrow,u_-(k)}\otimes \ket{\uparrow,u_-(k)}
\end{align}
Following the same logic as above, we have
\begin{align}
    \frac{1}{L^2}C_\text{yellow/orange} =  c_1[b] + c_1[b] = \begin{cases}
        -2, &\text{ orange}\\
        +2, &\text{ yellow}\\
        0, &\text{otherwise}
    \end{cases}
\end{align}
The compensating sign comes from the flavor-spin part of the wavefunction. Therefore, the yellow/orange regimes exhibit the IQHE but is a trivially gapped phase or metal otherwise. 

There is a subtlety in the preceding discussion. There are two distinct notions: singular points of the wavefunction and points in the parameter space where theory becomes gapless. In the case of $N_f=1$, these two notions coincide, i.e. the wavefunction was singular precisely where the band gap closes. But these two notions no longer coincide for $N_f=2$. There exist points $\mu_c$ where the Chern number does not jump since we do not encounter a singularity of the wavefunction but the gap closes as we dial $\mu\rightarrow\mu_c$ and we transition from a metallic to an insulating phase. This is possible because the wavefunction is independent of $\mu$, while the eigenvalues depend explicitly on $\mu$. Physically speaking, the Hall current in the metallic phase is negligible due to large ordinary electric current parallel to any infinitesimal applied electric field. Therefore, while the Chern number does not jump across a metal-insulator transition, it loses any physical meaning in a gapless phase. 

\subsubsection{Triplet mass} \label{App:Triplet}
With the same assumptions on the unit cell as before, we have 
\begin{align}
    H =  \sum_k\begin{pmatrix}
        \psi_1(k)\\ \psi_2(k)\end{pmatrix}^\dagger \Big[&\cH_x\Gamma^x+\cH_y\Gamma^y +\cH_z\sigma^z\otimes \sigma^z\no\\&+ \mu\; \sigma^z\otimes \mathbb{1}_2\Big]\begin{pmatrix}
        \psi_1(k)\\\psi_2(k)
    \end{pmatrix}
\end{align}
Define $\ket{u_\pm^{(s)}(k)} =\ket{ u_\pm(k, sm, sR)}$ with $s=\pm$ (corresponding to $s = \uparrow, \downarrow$) denoting the flavor $\sigma^z$ eigenvalues respectively. The eigenvectors that diagonalize the Hamiltonian are
\begin{align}
   \Big\{ \ket{\downarrow,u_\pm^{(-)}(k)},\ket{\uparrow, u_\pm^{(+)}(k)}\Big\}
\end{align}
which correspondingly have the eigenvalues $\Big\{-\mu \pm \sqrt{\cH_x^2 + \cH_y^2 + \cH_z^2}, +\mu \pm \sqrt{\cH_x^2 + \cH_y^2 + \cH_z^2}\Big\}$. As for the case of a singlet mass only, we have three distinct gapped regimes
\begin{align}
    \ket{0} = \begin{cases}
        \bigotimes_{k\in \cB} \ket{\downarrow,u_-^{(-)}(k)}\otimes \ket{\downarrow,u_+^{(-)}(k)}, &\mu\gg \sup|\cH|\\
        \bigotimes_{k\in \cB} \ket{\uparrow,u_-^{(+)}(k)}\otimes \ket{\downarrow,u_-^{(-)}(k)}, &|\mu|\ll \inf|\cH|\\
        \bigotimes_{k\in \cB} \ket{\uparrow,u_+^{(+)}(k)}\otimes \ket{\uparrow,u_-^{(+)}(k)}, &\mu\ll -\sup|\cH|
    \end{cases}
\end{align}
As before, we carry out the analysis of the occupation fraction as a function of the triplet mass and chemical potential. Since this plot is analogous to the preceding case, we summarize the results in Fig.~\ref{fig:PhaseDiag}. 

Consistently with our analysis of discrete symmetries on the lattice and the continuum, we find that the total Chern number is always vanishing. The interesting twist lies in the orange and yellow regions. Although the net Chern number vanishes, the Chern number for the spin-up/spin-down particles is not zero. In these regions, the Fock vacuum is
\begin{align}
    \ket{0}=\bigotimes_{k\in \cB} \ket{\uparrow,u_-^{(+)}(k)}\otimes \ket{\downarrow,u_-^{(-)}(k)}
\end{align}
Then, we see that $c_\uparrow = c_1[b]|_{+M} = -c_\downarrow = -c_1[b]|_{-M}$, where  $|_M$ denotes the value of the Chern number at mass $M$. Therefore, the net Chern number in the orange/yellow regions is zero, but the spin-up and spin-down Chern numbers are \textit{not}. In a insulating system with edges, this would give rise to the Quantum Spin Hall (QSH), where in the insulating bulk the Hall conductance and Hall currents vanish, but there are edge currents due to the equal and opposite non-zero Chern numbers of the spin-up and spin-down particles. 

\subsection{Time-reversal symmetry}
For the theory with equal masses and Wilson couplings, we simply have two copies of the $N_f=1$ Hamiltonian, which are both not $\cT$-invariant. Hence, this case breaks time-reversal symmetry, and indeed has phases with non-zero Chern numbers. 

On the other hand, the case with $(m_1, R_1) = -(m_2, R_2)$ can be shown to preserve $\cT$-symmetry. We start with the Hamiltonian in momentum space:
\begin{align}
    \cH = \cH_x\mathbb{1}_2&\otimes\sigma^x + \cH_y\mathbb{1}_2\otimes\sigma^y\no\\&+ \bigoplus_{j=1, 2}[m_j +R_j(2+\cos k_x+ \cos k_y)]\sigma^z
\end{align}

Acting with $\cT$: $(m_1, R_1)\rightarrow(-m_1, -R_1)$ and $(m_2, R_2)\rightarrow(-m_2, -R_2)$. This transformation leaves the theory invariant if and only if: 
\begin{align}
    m_1 &= -m_2 & R_1&=-R_2 &\psi_1\xleftrightarrow[]{\cT}\psi_2
\end{align}
This theory is, therefore, $\cT$-invariant provided we swap the two fermions and focus on equal and opposite masses and Wilson couplings. Hence, the Hamiltonian for the triplet case may be concisely written as
\begin{align}
    \cH = \cH_x\mathbb{1}_2\otimes\sigma^x + \cH_y\mathbb{1}_2\otimes\sigma^y + \cH_z\sigma^z\otimes\sigma^x.
\end{align}

\section{The many-body Chern number and perturbative robustness}
\label{sec:mbchern_robust}

To discuss the stability of the topological response under deformations like $H_E$, it is convenient to introduce the  {many-body Chern number} defined using  {twisted
boundary conditions} \cite{NiuThoulessWu1985,AvronSeilerSimon1985}.
On an $L\times L$ torus we impose twists
$\theta=(\theta_x,\theta_y)\in[0,2\pi)^2$,
so that translating a charged field around the $x$ (resp.\ $y$) cycle produces a phase
$e^{i\theta_x}$ (resp.\ $e^{i\theta_y}$). Equivalently, $\theta$ may be viewed as a
flat background $\mathrm{U}(1)$ gauge field threading the non-contractible cycles.
For each $\theta$, we denote the gauge-invariant ground state by
$\ket{\psi_0(\theta)}$, assumed to be nondegenerate and separated by a finite many-body gap for some $M$.

The associated Berry connection on the twist torus is
\begin{equation}
A_k(\theta) \;=\; i\,\braket{\psi_0(\theta)|\partial_{\theta_k}|\psi_0(\theta)},
\qquad k\in\{x,y\},
\end{equation}
with Berry curvature
\begin{equation}
F_{xy}(\theta) \;=\; \partial_{\theta_x}A_y(\theta)-\partial_{\theta_y}A_x(\theta).
\end{equation}
The  {many-body Chern number} is the first Chern number of the ground-state line bundle,
\begin{equation}
C \;=\; \frac{1}{2\pi}\int_{0}^{2\pi}\!\!d\theta_x \int_{0}^{2\pi}\!\!d\theta_y \; F_{xy}(\theta)
\;\in\;\integers,
\label{eq:mbchern_def}
\end{equation}
and in gapped phases it equals the quantized Hall response (in appropriate units)
\cite{NiuThoulessWu1985,AvronSeilerSimon1985}.
For later use (and to avoid gauge choices for $\ket{\psi_0}$), one may equivalently write
\cite{AvronSeilerSimon1983,AvronSeilerSimon1985}
\begin{equation}
C \;=\; \frac{i}{2\pi}\int_{0}^{2\pi}\!\!d\theta_x \int_{0}^{2\pi}\!\!d\theta_y\;
\mathrm{Tr}\!\Big(P(\theta)\,[\partial_{\theta_x}P(\theta),\partial_{\theta_y}P(\theta)]\Big),
\label{eq:chern_projector}
\end{equation}
where $P(\theta)=\ket{\psi_0(\theta)}\bra{\psi_0(\theta)}$ is the ground-state projector
(or, in the presence of a protected $r$-fold ground multiplet, the rank-$r$ spectral projector onto that
multiplet).

\subsection{Relation to momentum-space Chern number}
To understand the connection of $C$ with $c_1[b]$, we proceed by fixing the gauge $\ket{g_0} =\ket{\sfU_k=1}$ as we did in the preceding sections, but with twisted boundary conditions for the fermions, i.e.
\begin{align}
    \psi({r+L\hat{\ell}}) = e^{-i\theta_\ell}\psi(r) \text{ where } \ell = x,y.
\end{align}
Then the gauge-invariant ground state is:
\begin{align}
    \ket{\psi_0} = \amsmathbb{P}\ket{f_{g_0}^\theta, g_0}=\frac{1}{\sqrt{\abs{\amsmathbb{G}}}}\sum_{h\in\amsmathbb{G}}\ket{f_{h(g_0)}^\theta}\otimes\ket{h(g_0)}
\end{align}
where the state $\ket{f_g^\theta}$ is fermionic Fock ground state solved subject to the gauge configuration $g$ with twisted boundary conditions characterized by $\theta$. Observe that the gauge-ket does not depend on $\theta$. Hence, we obtain
\begin{align}
    \bra{\psi_0}\p_\theta\ket{\psi_0} = \frac{1}{\abs{\amsmathbb{G}}}\sum_{h\in \amsmathbb{G}}\bra{f_{h(g_0)}^\theta}\p_\theta\ket{f_{h(g_0)}^\theta}.
\end{align}

For each $h\in\amsmathbb{G}$, the (local) gauge transformation of fermions is
\begin{align}
h\,\psi(r)\,h^\dagger &= e^{i\alpha_h(r)}\psi(r),&\alpha_h(r+L\hat\ell)&=\alpha_h(r)\;\;(\ell=x,y),
\label{eq:Uh_def}
\end{align}
so that $h$ is  {independent of} $\theta$ and preserves the twisted boundary conditions
$\psi({r+L\hat\ell})=e^{-i\theta_\ell}\psi({r})$.
Gauge covariance of the Wilson--Dirac operator then implies that the fermion Hamiltonian in the background
$h(g_0)$ is unitarily equivalent to that in $g_0$ at the  {same} twist $\theta$,
\begin{equation}
H_f^{\theta}[\,h(g_0)\,] \;=\; h \, H_f^{\theta}[\,g_0\,] \, h^\dagger .
\label{eq:H_cov}
\end{equation}
Consequently, the many-body occupied-band Fock projector satisfies
\begin{equation}
P_-^{\theta}[\,h(g_0)\,] \;=\; h \, P_-^{\theta}[\,g_0\,] \, h^\dagger .
\label{eq:P_cov}
\end{equation}

Rather than working with the Berry connection (which depends on the phase convention for
$\ket{f_g^{\theta}}$), we use the gauge-invariant expression for the Berry curvature in terms of
$P_-^{\theta}[g]$,
\begin{equation}
F_{xy}[g](\theta)
\;=\;
i\,\mathrm{Tr}\!\left(
P_-^{\theta}[g]\;
\big[\partial_{\theta_x}P_-^{\theta}[g],\,\partial_{\theta_y}P_-^{\theta}[g]\big]
\right).
\label{eq:F_projector}
\end{equation}
Using \eqref{eq:P_cov} and cyclicity of the trace, we immediately obtain
\begin{equation}
F_{xy}[\,h(g_0)\,](\theta) \;=\; F_{xy}[\,g_0\,](\theta)
\qquad \forall\,h\in\amsmathbb{G}.
\label{eq:F_equal}
\end{equation}
In words:  {every gauge transform $h(g_0)$ yields the same Berry curvature as $g_0$ at fixed twist}.
Therefore all terms in the group average contribute identically at the curvature (and hence Chern-number)
level, and we may evaluate the many-body Chern number using any single representative of the orbit.

Since the fermionic ground state in each background is nondegenerate, then \eqref{eq:H_cov} implies
\begin{equation}
\ket{f_{h(g_0)}^{\theta}}
\;=\;
e^{i\chi_h(\theta)}\,h\,\ket{f_{g_0}^{\theta}},
\label{eq:state_phase}
\end{equation}
for some phase $\chi_h(\theta)$. Since $h$ is $\theta$-independent, the Berry connections
satisfy
\begin{equation}
i\braket{f_{h(g_0)}^{\theta}|\partial_{\theta_k}| f_{h(g_0)}^{\theta}}
=
i\braket{f_{g_0}^{\theta}|\partial_{\theta_k}| f_{g_0}^{\theta}}
+\partial_{\theta_k}\chi_h(\theta),
\end{equation}
so they can differ only by a pure gauge term. This term drops out of the curvature and hence cannot affect
the Chern number. In other words, up to pure-gauge terms:
\begin{align}
    \bra{\psi_0}\p_\theta\ket{\psi_0} =\bra{f_{g_0}^\theta}\p_\theta\ket{f_{g_0}^\theta}.
\end{align}
To calculate the above, we use the following Fourier expansion
\begin{align}
    \psi_r & = \frac{e^{-i\frac{1}{L}\theta\vdot r}}{\sqrt{L}}\sum_k e^{ik\vdot r}\psi_k
\end{align}
where the momenta satisfy
\begin{align}
   k_{n,m} = \left(k_x, k_y\right)  = \left(\frac{2\pi n}{L}, \frac{2\pi m}{L}\right)
\end{align}
where $n, m = 0, \dots, L-1$. Substituting this modified Fourier transform into the Hamiltonian, we obtain the modified Hamiltonian $\cH_\theta(k) = \cH(k-\theta)$ which gives:
\begin{align}
    \ket{f_{h(g_0)}^\theta} = \bigotimes_{n, m}\ket{u_-^{h(g_0)}(\tfrac{2\pi n}{L} - \tfrac{\theta_x}{L}, \tfrac{2\pi m}{L}-\tfrac{\theta_y}{L})}
\end{align}
The above expression may be further simplified using the definition of the state $\psi$:
\begin{align}\label{ChSimp}
    C = \frac{i}{2\pi} \sum_{n, m = 0}^{L-1}\oint_\gamma d\theta_i\; \bra{u_-(k_{n,m}-\tfrac{\theta}{L})} \partial_{\theta_i} \ket{u_-(k_{n,m}-\tfrac{\theta}{L})}
\end{align}
The evaluation of these integrals requires some care as the singularities in the wavefunctions occur at different values for each term in the sum. Let us evaluate the integral for fixed $(n,m)$, and then sum over $n$ and $m$. We notice that there is a singularity whenever
\begin{align}
    \frac{1}{L}(\theta_x, \theta_y) = \left(\frac{2\pi n}{L} - p\pi, \frac{2\pi m}{L} - q\pi\right)
\end{align}
for any $p, q\in\{0,1\}$. To evaluate the integral, it is convenient to change linear change of variables 
\begin{align}
    \theta_{n,m}=(\theta_{xn},\theta_{ym}) &= \left(\frac{2\pi n}{L}-\frac{\theta_x}{L}, \frac{2\pi m}{L} - \frac{\theta_{y}}{L}\right)
\end{align}
This changes the summand in \eqref{ChSimp} to 
\begin{align}
    \oint_{\gamma_{n,m}}d\theta_{n,m} \bra{u_-(\theta_{n,m})}\partial_{\theta_{n,m}} \ket{u_-(\theta_{n,m})}
\end{align}
Using the periodicity of the integrand, we note that each term above evaluates to $c_1[b]$, which gives:
\begin{align}\label{Cnet}
    C = c_1[b]
\end{align}Hence, the many-body Chern number at weak coupling is simply $c_1[b]$.

\subsection{Robustness under deformations}\label{sec:Robustness}

As before, we split the Hamiltonian at fixed gauge coupling $e^2>0$ as $H = +H'$
\begin{equation}
H \;=\; H_f + H_m - \frac{1}{e^2}H_B,
\qquad
H' \;=\; e^2 H_E,
\end{equation}
where $H_E$ and $H_B$ carry no explicit powers of $e^2$.
To make the robustness statement precise, we introduce an interpolation parameter $s\in[0,1]$ and define
\begin{equation}
H^{(s)}(\theta) \coloneqq H(\theta) + s\,H'
= H_f(\theta) + H_m - \frac{1}{e^2}H_B + s\,e^2 H_E .
\label{eq:Hs_interp}
\end{equation}
Let $P^{(s)}(\theta)$ denote the ground-state spectral projector of $H^{(s)}(\theta)$
in the physical Hilbert space. If there is a  {uniform many-body gap} along the entire path,
\begin{equation}
\Delta_{\min}\;:=\;\inf_{s\in[0,1]}\;\inf_{\theta\in\mathbb{T}^2}\;
\Big(E_{r}(\theta,s)-E_{0}(\theta,s)\Big) \;>\; 0,
\label{eq:uniform_gap}
\end{equation}
then the ground-state (or ground-subspace) spectral projector $P^{(s)}(\theta)$
varies smoothly with $(s,\theta)$. A convenient formulation of this
smoothness is provided by  {quasi-adiabatic continuation}: as long as the gap
\eqref{eq:uniform_gap} remains open, one can construct a continuous family of
(quasi-local) unitaries $U(s)$ that transports the ground-state subspace along the
path, so that
\begin{equation}
P^{(s)}(\theta) \;=\; U(s;\theta)\,P^{(0)}(\theta)\,U^\dagger(s;\theta),
\end{equation}
which in particular implies smooth dependence of $P^{(s)}(\theta)$ on the
parameters \cite{HastingsWen2005,HastingsMichalakis2015}. In this case the Chern number computed from \eqref{eq:mbchern_def}
is  {constant} as a function of $s$:
\begin{equation}
C(s) \equiv C\!\left[H^{(s)}\right]
\end{equation}
is independent of $s$ as long as \eqref{eq:uniform_gap} holds. Equivalently, the integer topological invariant cannot change under a continuous deformation that
preserves the gap; any change of $C$ requires a gap closing (or a change in the rank of the ground
projector) somewhere on the twist torus \cite{AvronSeilerSimon1983,AvronSeilerSimon1985}. Defining the  {full} invariant to be the Chern number of the complete Hamiltonian,
\begin{equation}
C_{\mathrm{full}} \;:=\; C\!\left[H^{(1)}\right],
\end{equation}
the above implies the desired equality
\begin{equation}
C_{\mathrm{full}}
\;=\; C\!\left[H^{(1)}\right]
\;=\; C\!\left[H^{(0)}\right],
\label{eq:Cfull_equals_C0}
\end{equation}
i.e.\ turning on the electric term $e^2H_E$ from $s=0$ to $s=1$ does not change the many-body Chern number
provided the many-body gap remains open for all twists. For numerical evaluation on a discrete twist grid, we use a gauge-invariant lattice formula for the Berry curvature, following Ref.~\cite{FukuiHatsugaiSuzuki2005}. 

\section{Probing Topological Phases with Current Correlators}\label{sec:CurrentCorr}
\subsection{The case of $N_f=1$ flavor}
Consider the gauge-current operator with $k = x, y$
\begin{align}
    J_k(r) &= \frac{\d }{\d\log \sfU_k(r)}H[\sfU] \no\\&= \psi(r)^\dagger M_k\psi(r+\hat{k}) \sfU_k - \frac{1}{e^2}\sum_{(r,k)\in \Box} \sfU_\Box +h.c.
\end{align}
which is the lattice counterpart of the continuum Chern-Simons current. Note that the second term comes from the $H_B$ and the sum runs over those plaquettes that contain the link $k$ with base point $r$. Since the plaquette contribution is simply an additive constant (for a fixed $e^2$) in the trivial-flux sector, we may simply drop in the following calculation. This indicates that fermions and gauge fields decouple in this weakly coupled sector.

Since we are working in the gauge choice of Sec.~\ref{sec:TrivialFlux}, we may consider the correlation function of $J_k$ with $\sfU_k=1$. To compute this correlator, it is convenient to write the Fourier transform as:
\begin{align}
    \psi(r) = \frac{1}{L} \sum_{k\in\cB}e^{ik\vdot r}\psi(k) = \frac{1}{L} \sum_{k\in\cB}e^{ik\vdot r}\cU(k)^{-1}\tilde{\psi}_k
\end{align}
where $\tilde\psi$ is the basis in which the free fermionic Hamiltonian is diagonal. In this basis, the current operator is
\begin{align}
    J_k(r) = \sum_{p, q\in\cB}\tilde{\psi}_p^\dagger \frac{e^{-ip\vdot r}}{L} [\cU(p) M_k\cU(q)^\dagger ]\frac{e^{iq\vdot(r+\hat{k})}}{L}\tilde{\psi}_q+h.c.
\end{align}
We wish to calculate the expected value of this operator in the Fock vacuum, given by
\begin{align}
    \ket{0} = \bigotimes_{k\in\cB}\psi_{k-}^\dagger \ket{\emptyset}
\end{align}
To achieve this, we compute the momentum-space two-point function with $s, s'\in \{\pm\}$:
\begin{align}
    \bra{0}\psi_s(p)^\dagger \psi_{s'}(q)\ket{0} \label{2pf}
\end{align}
Consider
\begin{align}
    \psi_{s'}(q)\ket{0} = \{\psi_{s'}(q),\psi^\dagger_-(q)\}\bigotimes_{q'\neq q}\psi_-(k)^\dagger \ket{\emptyset}\no\\ = \d_{s',-}\bigotimes_{q'\neq q}\psi_-(k)^\dagger \ket{\emptyset}
\end{align}
\begin{figure*}
    \centering
    \includegraphics[width=0.4\linewidth]{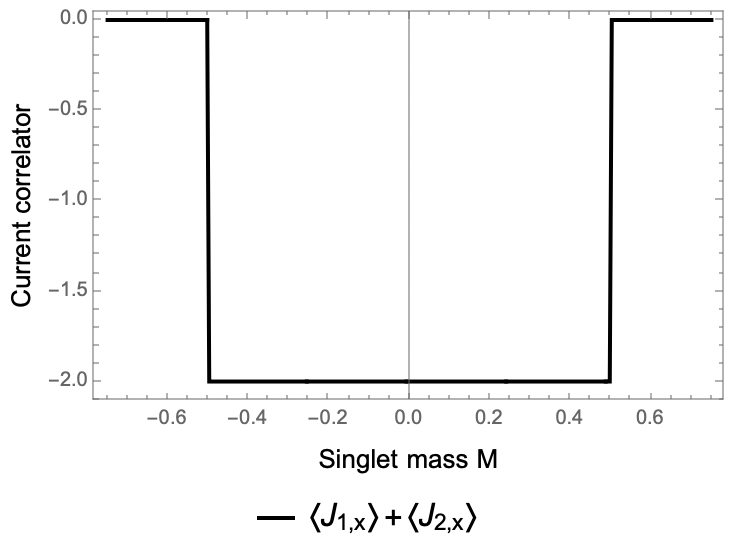} 
    \hspace{3em}
    \includegraphics[width=0.4\linewidth]{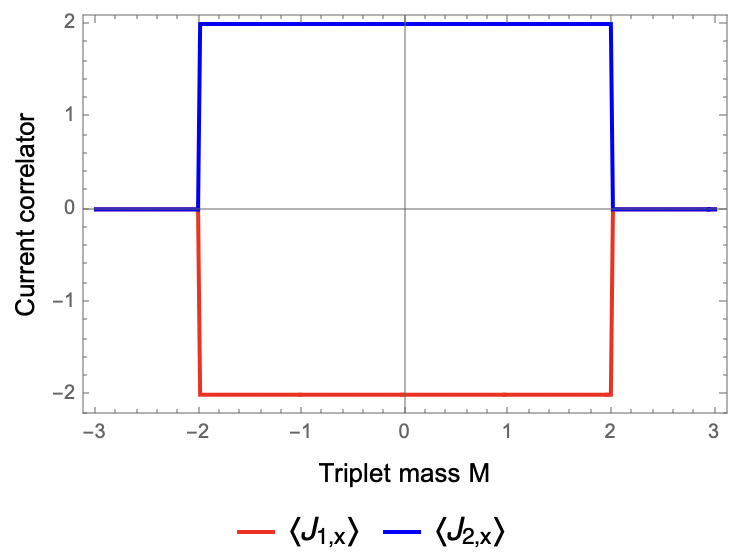}
    \caption{The analytical current one-point functions for both the singlet and triplet cases, which are non-trivial in the IQH and QSH phases respectively. }
    \label{fig:current}
\end{figure*}
Likewise, it follows that 
\begin{align}
    \bra{0}\psi_s(p)^\dagger  = \d_{s,-}\bigotimes_{p'\ne p}\bra{\emptyset}\psi_-(p')
\end{align}
It is clear that whenever $p\ne q$, the momentum-space two-point function \eqref{2pf} vanishes by the definition of the empty-state $\ket{\emptyset}$. Putting it all together
\begin{align}
    \bra{0}J_k(r)\ket{0} = \sum_{p\in \cB} \frac{ [\cU(p) M_k\cU(p)^\dagger ]^{--}{e^{ip\vdot\hat{k}}}}{L^2} + h.c.
\end{align}
The above expression is trivial to evaluate with the aid of copmuter algebra, but not particularly illuminating to display here for general $L$. Importantly, the above expression has the property that it is non-zero precisely when we are in a topological phase! For clarity, we spell out the case when $L=2$, which gives four contributions -- see Table.~\ref{tab:Contributions}. This implies:
\begin{align}
    \bra{0}J_x(r)\ket{0} &=(-)^{\frac{1+\text{sign}(M+2)}{2}}\frac{1}{4}\abs{1+\frac{M+2}{\abs{M+2}}}\no\\&+ (-)^{\frac{1-\text{sign}(M-2)}{2}}\frac{1}{4}\abs{1+\frac{M-2}{\abs{M-2}}}\no\\
    &= \begin{cases}
        -1,& \abs{M}<2 \text{ (topological)}\\
        \;\;\,0, & \abs{M}>2 \text{ (trivial)}\\
    \end{cases}
\end{align}
Hence, $J_k$ serves as a robust marker of topological phases --   exactly as the continuum Chern-Simons current obtained after integrating out a massive Wilson fermion!

\begin{table}
    \centering
    \begin{tabular}{c|c}
        $(k_x, k_y)$& Contribution to $\bra{0}J_x\ket{0}$\\
        \hline
        \hline
        $(0,0)$&  $(-)^{\frac{1+\text{sign}(M+2)}{2}}\frac{1}{4}\abs{1+\frac{M+2}{\abs{M+2}}}$\\
        \hline
         $(0, \pi)$ & $-\frac{1}{4}\abs{1+\frac{M}{\abs{M}}}$\\
         \hline
         $(\pi,0)$ & $+\frac{1}{4}\abs{1+\frac{M}{\abs{M}}}$\\
         \hline
         $(\pi, \pi)$ &$(-)^{\frac{1-\text{sign}(M-2)}{2}}\frac{1}{4}\abs{1+\frac{M-2}{\abs{M-2}}}$
    \end{tabular}
    \caption{Summary of contributions to $\braket{J_x}$ for $L=2$.}
    \label{tab:Contributions}
\end{table}

\subsection{The case of $N_f=2$ flavors}
The fermionic contribution to the gauge current operator is
\begin{align}
    J_k^a(r) &= \frac{\d H_a[\sfU]}{\d \log \sfU_k(r)}= \psi_{a}(r)^\dagger M_k \sfU_k(r)\psi_a(r+\hat{k}) +h.c.
\end{align}
For the unperturbed problem without the $H_E$ term, we may go to the gauge in which $\sfU_k=1$ in the trivial-flux sector, as in Sec.~\ref{sec:TrivialFlux}. Supposing that 
\begin{align}
    \cU(k) = \begin{psmallmatrix}\cU_{11}(k) & \cU_{21}(k)\\\cU_{12}(k) & \cU_{22}(k)\end{psmallmatrix}
\end{align}
is the $2N_f\times 2N_f$ matrix that diagonalizes the Hamiltonian such that $\psi_a(k) = \cV_{a,s}(k)\Psi_s(k)$, where $\Psi_s = (\tilde{\psi}_{1,s}, \tilde{\psi}_{2,s})^\top$ is a $2$-component spinor with $s = \pm$ and $\cV = \cU^{-1}$ and the four $2\times 2$ blocks of $\cV$ are labeled as $\cV_{a,s}$. Then it follows that 
\begin{align}
    \psi_a(r) = \frac{1}{L}\sum_{k\in\cB}e^{ik\vdot r}\psi_a(k) = \frac{1}{L}\sum_{k\in\cB}e^{ik\vdot r}\cV_{a,s}(k)\Psi_s(k)
\end{align}
\begin{figure*}
    \centering
    \includegraphics[width=0.45\linewidth]{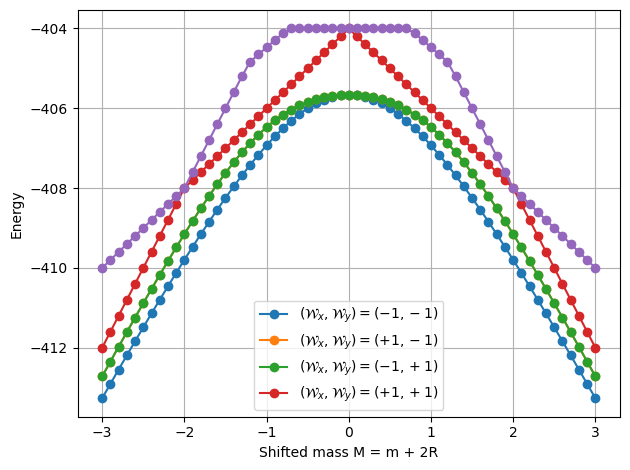}
    \hspace{3em}\includegraphics[width=0.45\linewidth]{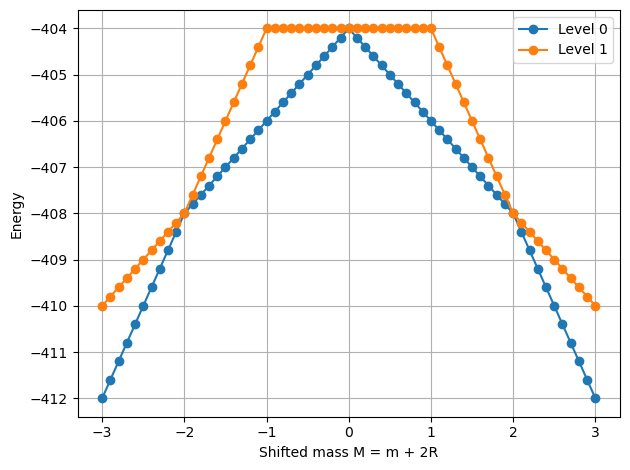}
    \caption{The lowest five energies versus $M$ for the $N_f=1$ theory before projecting onto the trivial-flux sector (left), and the level crossing after projecting onto the trivial-flux sector (right) with $(\cW_x, \cW_y) = (1,1)$ via $\prod_k\frac{1+\cW_k}{2}$. The results were obtained via ED on a $2\times 2$ lattice with $\integers_2$ gauge fields and $e^2 = 0.01$. We emphasize that the left figure did \textit{not} require a projection into the approximate $(\cW_x, \cW_y)$ superselection sectors, but rather were found to naturally lie in the four sectors labeled as $(\pm1, \pm1)$}
    \label{fig:LevelCrossing}
\end{figure*}
Then:
\begin{align}
    J_k^a= \sum_{p, q\in\cB}\frac{{\Psi}_s(p)^\dagger  \cV_{as}(p)^\dagger M_k\cV_{as'}(q){\Psi}_{s'}(q) {e^{iq\vdot(r+\hat{k}) - ip\vdot r}}}{L^2}\no\\+h.c.
\end{align}
where $s$ and $s'$ are implicitly summed over. As before, evaluating the one-point function $\bra{0}J_x\ket{0}$ boils down to knowing momentum-space two-point function
\begin{align}
    \bra{0}\tilde{\psi}_{cs}(p)^\dagger\tilde{\psi}_{bs'}(q)\ket{0} = \delta_{bc} \delta_{s,-}\d_{s', -}\d(p-q)
\end{align}
Hence:
\begin{align}
    \bra{0}J_k^a\ket{0}&= \frac{1}{L^2}\sum_{p\in\cB} \Tr[\cV_{a-}(p)^\dagger M_k\cV_{a-}(q)] {e^{ip\vdot\hat{k}}}+h.c.
\end{align}
It is now straightforward to plug in expressions for the matrices $\cV$ and $M_k$ to obtain the correlator, which has been plotted in Fig.~\ref{fig:current}. In perfect agreement with the results for the $N_f=1$ case as well as the computation of the Chern number, we find:
\begin{align}
    \begin{cases}
        \text{singlet: } J_k = J_k^1 + J_k^2 = \begin{cases}
            -2,&\text{ when }\abs{M}<2\text{ (top.)}\\
            \;\;\;0,&\text{ otherwise (trivial)}
        \end{cases}\\
        \text{triplet: } J_k^1 = -J_k^2 = \begin{cases}
            -1,&\text{ when }\abs{M}<2 \text{ (top.)}\\
            \;\;\;0,&\text{ otherwise (trivial)}
        \end{cases}
    \end{cases} 
\end{align}

In summary, the current correlator at weak coupling is a robust $\integers_2$ marker for topological phases.

\section{A path to Wilson fermion quantum simulations of QED$_3$}\label{sec:Numerics}

{So far, we have shown that lattice Hamiltonian theories with Wilson fermions capture a variety of topological phases while staggered fermions fail to do so. However, staggered fermions have received a lot of attention as far as implementation on quantum computing platforms is concerned, while similar experiments and quantum simulations with Wilson fermions are not as widespread. In this section, we provide numerical results obtained via exact diagonalization (ED) on $2\times2$ systems, with and without gauge fields, which serve as a starting point for a deeper study of quantum simulations with Wilson fermions. This also provides an independent check on the results developed so far. }

When we include gauge fields, there are a number of considerations. First, the gauge fields must be truncated from $\U(1)$ to $\integers_N$. Second, the physical states that carry both fermionic and gauge labels must satisfy Gauss' law and be gauge-invariant. Third, the symmetries imposed prior to introducing the gauge fields must be compatible with the gauge symmetry, i.e. the Gauss-law projector and the symmetry-generators must share simultaneous eigenstates. We have explicitly shown this for the $\integers_N$ and $\U(1)$ cases with {$\cW_k$ flux-symmetry} in Sec.~\ref{sec:FluxSym}. Since our theoretical arguments indicate that the lowest $\cW_k$-invariant state is independent of the gauge field values, we must be able to see the topological transitions from ED even for $N=2$ as a proof of concept. This offers a possibility to realize such topological phases in near-term quantum devices via spin representation for the truncated gauge group, such as superconducting devices~\cite{Cochran:2024rwe}, Rydberg atoms~\cite{gonzalez2024observation}, {trapped} ion~\cite{Meth:2023wzd}, dipolar molecules~\cite{luo2020framework} and fermion-pair registers~\cite{sun2023quantum}.

\begin{figure}[b]
    \centering
    \includegraphics[width=\linewidth]{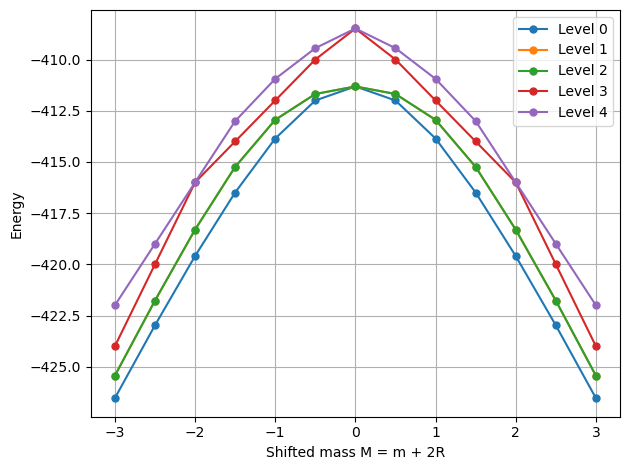}
    \caption{The lowest five energies versus $M$ for the $N_f=2$ theory with $\integers_2$ gauge fields and $e^2 = 0.01$. The various energy levels live in distinct approximate superselection sectors labeled by $(\cW_x, \cW_y)$ at weak coupling. The results were obtained via ED on a $2\times 2$ lattice. No projection into the superselection sectors was performed here.}
    \label{fig:LevelCrossing2}
\end{figure}

\begin{figure}[b]
    \centering
    \includegraphics[width=0.9\linewidth]{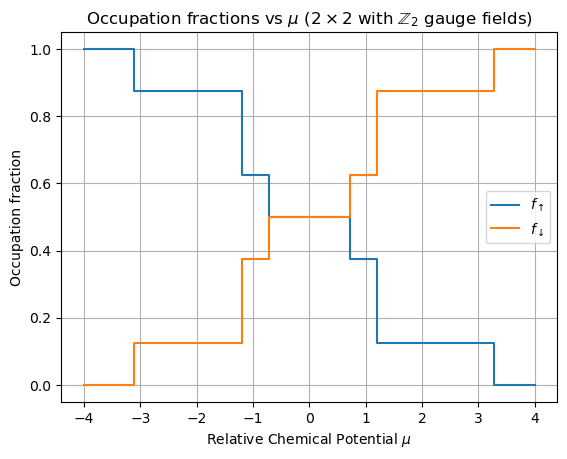}
    \caption{The average occupation fraction $f_a$ versus the relative chemical potential $\mu$ for the $N_f=2$ theory for a fixed gauge coupling $e^2 = 0.01$ in the lowest level with $(\cW_x, \cW_y)\approx (1,1)$; we did \textit{not} project into this subspace.}
    \label{fig:finiteDensity}
\end{figure}

In this section, we provide an algorithmic summary of our numerical simulations carried out via exact diagonalization (ED). Let us outline the key results before we dive into the algorithmic details. Fig.~\ref{fig:LevelCrossing} displays the results of ED for the $N_f=1$ case. The lowest four levels live in the four distinct approximate superselection sectors that exist at weak coupling. Of key interest here is the $(1,1)$ sector, which exhibits level crossing and corresponding topological phase transitions at $M = 0, \pm 2$. Fig.~\ref{fig:LevelCrossing2} displays the analogous result for the $N_f=2$ case. The finite density phase transitions are verified in the $N_f=2$ theory in Fig.~\ref{fig:finiteDensity}. Next, we summarize parts of the algorithm that are common to both the $N_f=1$ and $N_f=2$ cases (i.e. the gauge sector), and then we handle the parts which are different one-by-one (i.e. the fermionic sectors). Following this, we discuss the numerical calculations of a variety of observables, including the current one-point function and the many-body Chern number, and the corrections to their expectation-values as a function of the gauge coupling. 

\subsection{The gauge sector}
We work on a periodic $2\times 2$ lattice with periodic boundary conditions $\amsmathbb{Z}_2$ gauge fields on the links. The gauge Hilbert space is
\[
\mathcal{H}_G = (\amsmathbb{C}^2)^{\otimes 8},\qquad \dim\mathcal{H}_G = 2^8=256,
\]
with computational basis $\{|g\rangle: g\in\{0,1\}^8\}$. Let $\sfG(x)$ denote the local $\amsmathbb{Z}_2$ Gauss' law operator at site $x$, acting by a star flip on the four incident links and a matter-parity phase,
\begin{equation}
\sfG(x)\,|f,g\rangle
=
(-1)^{n_x(f)}\,|f;\,g\oplus \cM_x\rangle,
\label{eq:gauss_action_general}
\end{equation}
where $\cM_x$ is the star mask, $\oplus$ is XOR, and $n_x(f)$ is the occupation number at $x$ (defined below for each $N_f$). Rather than forming the Gauss' law projector $\amsmathbb{P}$ as a matrix, we construct an isometry $V$ whose adjoint implements the projector, i.e. $V^\dagger V = \amsmathbb{P}/\sqrt{|\amsmathbb{G}^{(2)}|}$ and $V V^\dagger =\mathbb{1}_{\rm phys}$.
Concretely, for each gauge orbit $\mathcal{O}=\{g\oplus \cM_h:\,h\in\amsmathbb{G}^{(2)}\}$ and fixed matter configuration, one averages over $\amsmathbb{G}^{(2)}$
with the phases accumulated from \eqref{eq:gauss_action_general}.
Equivalently, $V^\dagger$ embeds a physical state into the full space by distributing amplitudes over each orbit with the appropriate $\pm 1$ signs,
and $V$ maps back by orbit-averaging. At fixed coupling $e^2$ we write
\begin{equation}
H(m)= (m+2R)\,H_{m} + H_f \;-\;\frac{1}{e^2}H_{B} \;+\; e^2 H_{E},
\label{eq:H_decomp}
\end{equation}
where $H_{B}$ is diagonal and equals the plaquette-flux sum $\sum_{\square}\prod_{\ell\in\partial\square} Z_\ell$,
and $H_E=\sum_{\ell} X_\ell$ flips individual links. The kinetic term is the Wilson nearest-neighbor hopping including the link signs $Z_\ell(g)=\pm 1$.
The physical Hamiltonian is always obtained by projection,
\begin{equation}
H^{\rm phys}(m) \;=\; V\,H(m)\,V^\dagger.
\label{eq:H_phys_def}
\end{equation}
Next, we address the fermionic sectors for $N_f=1, 2$ one-by-one.

\subsection{$N_f=1$ QED$_3$}

For $N_f=1$, matter consists of a two-component Wilson fermion on each site, i.e.\ $2L^2=8$ modes.
We restrict to half filling $\sum_{\alpha=1}^8 n_\alpha=4$ and define
\begin{align}
    \mathcal{H}_F^{(1)}&=\mathrm{span}\{|f\rangle:\ f\in\{0,1\}^8,\ |f|=4\},
\end{align}
where $\dim\mathcal{H}_F^{(1)}=\binom{8}{4}=70$ and impose Gauss' law. Thus, $\mathcal{H}_{\rm phys}^{(1)}=\mathcal{H}_F^{(1)}\otimes\mathcal{H}_G$ has dimension $70\times 32=2240$. In \eqref{eq:gauss_action_general} the parity is the on-site occupation parity of the single flavor,
\[
n_x(f) \;=\; \Big(\sum_{s=0}^1 n_{x,s}(f)\Big)\bmod 2,
\]
i.e. $(-1)^{n_x(f)}=-1$ iff the site has odd fermion number. 

\subsection{$N_f=2$ QED$_3$}
For $N_f=2$, we take two identical flavors (labeled $\uparrow,\downarrow$), each with $8$ modes, and work in the fixed-number sector $(N_\uparrow,N_\downarrow)=(4,4)$.
Let $\mathcal{H}_8^{(4)}$ denote the single-flavor $8$-mode Hamming-weight-$4$ subspace, $\dim\mathcal{H}_8^{(4)}=70$.
Then
\[
\mathcal{H}_F^{(2)}=\mathcal{H}_8^{(4)}\otimes\mathcal{H}_8^{(4)},\qquad \dim\mathcal{H}_F^{(2)}=70^2=4900,
\]
and $\mathcal{H}_{\rm phys}^{(2)}=\mathcal{H}_F^{(2)}\otimes\mathcal{H}_G$. In \eqref{eq:gauss_action_general} the parity is the  {total} on-site occupation parity summed over both flavors and both spinor components,
\[
Q_x(f_\uparrow,f_\downarrow)
\;=\;
\Big(\sum_{s=0}^1 n_{x,s}(f_\uparrow)+\sum_{s=0}^1 n_{x,s}(f_\downarrow)\Big)\bmod 2.
\]

Since the group acts only on the gauge register, the $256$ gauge configurations decompose into $256/8=32$ orbits of size $8$.
Accordingly the projected representation is naturally organized as amplitudes on $\mathcal{H}_F^{(2)}\otimes\amsmathbb{C}^{32}$ and has dimension
\[
\dim\mathcal{H}_{\rm phys}^{(2)} = 4900\times 32 = 156800.
\]
We do not explicitly assemble $H^{\rm phys}(m)$.
Instead, we implement its action by the three-step map $x \xmapsto{V^\dagger} y \xmapsto{H(m)} z \xmapsto{V} H^{\rm phys}(m)x$, where $x$ and $H^{\rm phys}(m)x$ are stored in the orbit basis and $y,z$ in the full gauge basis. Low-lying eigenvalues are then obtained by diagonalization.

\begin{figure}[b]
    \centering
    \includegraphics[width=\linewidth]{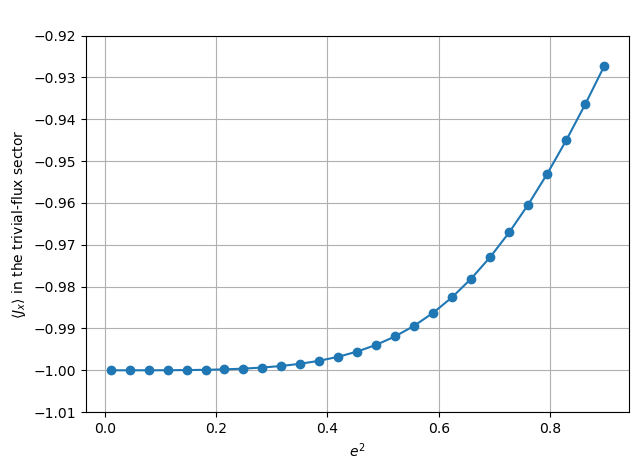}
    \caption{Corrections to the current one-point function at a fixed shifted mass $M =0.5$ in the $N_f=1$ theory.}
    \label{fig:NonPet}
\end{figure}
\subsection{Current Correlators}
Using the ED algorithm, it is straightforward to evaluate the current one-point function in the presence of gauge interactions to verify the analytical predictions of Sec.~\ref{sec:CurrentCorr}. The results of Fig.~\ref{fig:JxChern} clearly indicate that in the trivial-flux sector with $(\cW_x, \cW_y) = (1,1)$, the fermionic contribution to the current one-point function $\braket{J_x^{(f)}}\ne0$ only when the theory is in a topologically non-trivial phase. Otherwise, it must vanish. Interestingly, we note that the current one-point function in other superselection sectors is smooth and does not exhibit any sharp jumps, indicating an absence of phase transitions in all sectors except $(\cW_x, \cW_y) = (1,1)$. We emphasize that the results are \textit{not} projected down into the superselection sectors but are found by evaluating $\cW_x$ and $\cW_y$ for the lowest states in the energy spectrum. As anticipated, we find there are non-perturbative corrections to the wave-function and hence observables, including $\braket{J_x}$, which are suppressed at weak coupling.  Fig.~\ref{fig:NonPet} displays the corrections to $\braket{J_x}$ vs. $e^2$.

\begin{figure}[b]
    \centering
    \includegraphics[width=\linewidth]{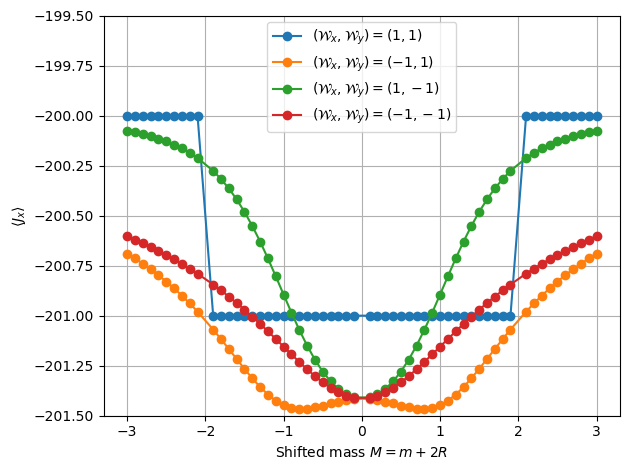}
    
    \caption{The current one-point function $\braket{J_x}$ versus $M$ for a fixed gauge coupling $e^2 = 0.01$, i.e. at weak coupling. The blue curve corresponds the expectation in the (approximate) $(1,1)$ trivial-flux sector in the $N_f=1$ theory.}
    \label{fig:JxChern}
\end{figure}
\begin{figure}[b]
    \centering
    \includegraphics[width=\linewidth]{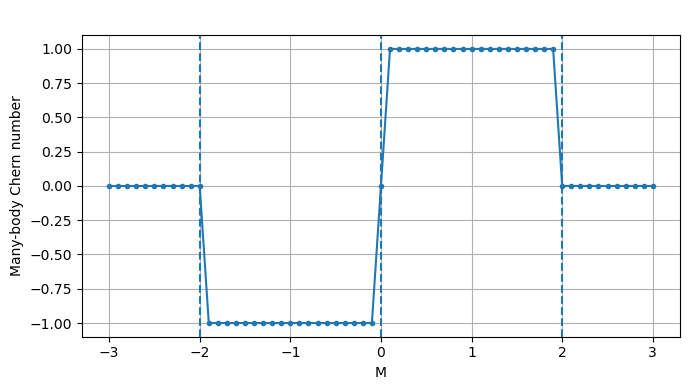}

    \caption{The many-body Chern number $C$ for a fixed gauge coupling $e^2 = 0.01$, i.e. at weak coupling computed in the (approximately) trivial-flux state with $(\cW_x, \cW_y)\approx(1,1)$ in the $N_f=1$ theory.}
    \label{fig:JxChern}
\end{figure}

\subsection{Many-body Chern number from twists}
\label{sec:chern_fhs}
We compute the many-body Chern number of a chosen eigenstate by imposing $\U(1)$ boundary twists
$\theta=(\theta_x,\theta_y)\in[0,2\pi)^2$ and evaluating the Berry curvature on the twist torus
using the Fukui--Hatsugai--Suzuki (FHS) lattice discretization. The twists enter only through the boundary-wrapping hops of the Wilson kinetic term.
Introduce directed ``cut'' operators $e^{i\theta_x}$ and $e^{i\theta_y}$ that contain precisely those hoppings crossing the $x$- and
$y$-boundaries, and define
\begin{align}
H_x(\theta_x) &=
H_x^{\mathrm{bulk}}
+
e^{\,i s\,\theta_x}e^{i\theta_x}
+
e^{-i s\,\theta_x}e^{-i\theta_x},
\nonumber\\
H_y(\theta_y) &=
H_y^{\mathrm{bulk}}
+
e^{\,i s\,\theta_y}e^{i\theta_y}
+
e^{-i s\,\theta_y}e^{-i\theta_y},
\label{eq:twist_cut_prd}
\end{align}
with $s=\pm1$ a fixed phase convention. Then
$H_f(\theta)=H_x(\theta_x)+H_y(\theta_y)$. We discretize the twist torus by $\theta_\mu^{(i)}=2\pi i/N_\theta$ for $i=0,\ldots,N_\theta-1$.
At each grid point $\theta_{ij}=(\theta_x^{(i)},\theta_y^{(j)})$ we diagonalize
$H^{\mathrm{phys}}(\theta_{ij})$ in a small energy window and select the target branch by continuity: starting from a reference eigenstate at $(0,0)$, we choose at neighboring points the eigenvector with maximal overlap
with the previously selected state. This overlap criterion robustly tracks the same eigenstate across the grid
away from gap closings.
From normalized states $\{\ket{\psi_{ij}}\}$ we form the $\U(1)$ link variables
\begin{equation}
U_x(i,j)=\frac{\braket{\psi_{ij}|{\psi_{i+1,j}}}}{\bigl|\braket{\psi_{ij}|{\psi_{i+1,j}}}\bigr|},
\qquad
U_y(i,j)=\frac{\braket{\psi_{ij}|{\psi_{i,j+1}}}}{\bigl|\braket{\psi_{ij}|{\psi_{i,j+1}}}\bigr|},
\label{eq:UxUy_prd}
\end{equation}
with periodic identification of indices modulo $N_\theta$.
The field strength on each plaquette of the twist grid is
\begin{equation}
F_{ij}=
\arg\left[
U_x(i,j)U_y(i+1,j)
U_x(i,j+1)^{-1}U_y(i,j)^{-1}
\right],
\label{eq:Fij_prd}
\end{equation}
and the FHS estimator of the many-body Chern number is
\begin{equation}
C=\frac{1}{2\pi}\sum_{i,j=0}^{N_\theta-1}F_{ij}.
\label{eq:Chern_prd}
\end{equation}
In gapped phases $C$ converges rapidly to an integer as $N_\theta$ is increased; deviations occur near gap closings,
where the tracked eigenstate can become ill-defined on the twist torus. As a consistency check, in regimes where the two flavors contribute additively, we check that
$C_{N_f=2}=2\,C_{N_f=1}$ at the same $(m,e^2)$. The results are summarized in Fig.~\ref{fig:JxChern}.
\begin{figure}[b]
    \centering
    \includegraphics[width=\linewidth]{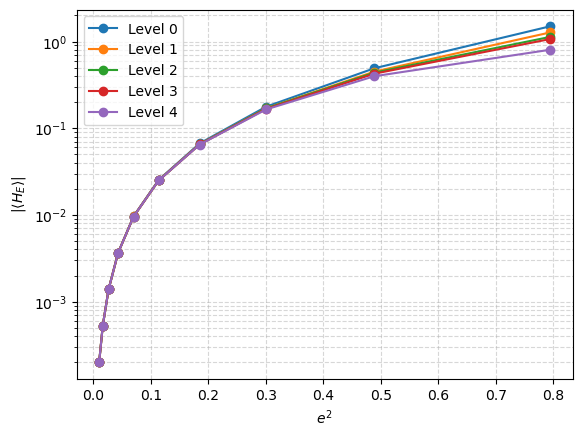}
    \caption{$\abs{\braket{H_E}}$ vs. $e^2$ at $M=-1$ in the lowest five eigenstates for truncated gauge group $\integers_2$.}
    \label{fig:Elec}
\end{figure}
\subsection{Effect of the electric term on the spectrum}
\begin{figure*}
    \centering
    \includegraphics[width=0.45\linewidth]{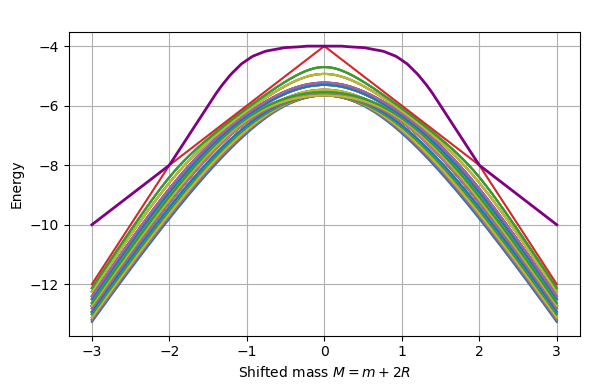}
    \hspace{3em}\includegraphics[width=0.45\linewidth]{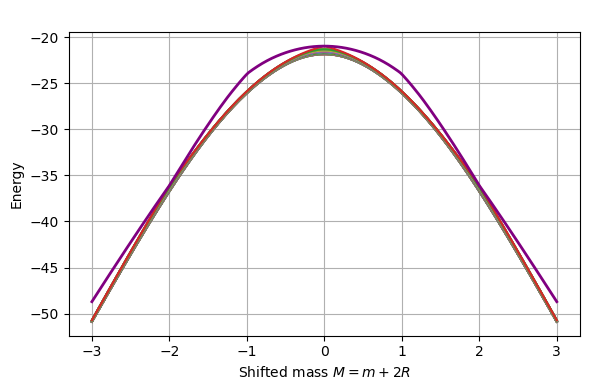}
    \caption{The lowest $N^2+1$ energy levels with $N=8$ for $2\times 2$ (left) and $4\times 4$ (right) lattices respectively.}
    \label{fig:FiniteSizeEffectsSpectrum}
\end{figure*}
Although the $e^2$-\textit{independent} electric energy operator $H_E = \sum_{k, r}\sfE_k(r)^2$ is suppressed at weak-coupling, we emphasize that it contributes small off-diagonal matrix elements that modify the energies. We study the contribution $\abs{\braket{H_E}}$ of the electric term to the energies.
These off-diagonal corrections to the energies do \textit{not} modify the topological phase structure, which is robust to such smooth deformations, thanks to the adiabatic continuity argument presented in Sec.~\ref{sec:Robustness}. Nonetheless, Fig.~\ref{fig:Elec} is instructive because it demonstrates two things. First, it shows that the small-ness of $H_E$ is \textit{not} trivially due to the multiplicative pre-factor $e^2/2$ being small; the $e^2$-\textit{independent} operator expectation $\braket{H_E}$ is itself suppressed at weak-coupling. Second, it shows that the gauge-kinetic terms at weak-coupling do not merely modify the energies by a trivial shift, but rather also contribute off-diagonal terms that induce non-trivial changes in the spectrum as $e^2$ is dialed up. A detailed analysis of such strongly-coupled phases is expected to reveal signatures of confinement and, hence, is of great interest. We leave this for future work.

\section{Analysis of Truncation and Finite-Size Effects}\label{sec:TruncationFiniteSize}

In this section, we analyze the effects of truncation and finite-size effects using a complementary approach to the exact diagonalization presented in the preceding section. In particular, we emphasize that we do \textit{not} use the exact-diagonalization (ED) algorithm, which requires us to construct an exponentially large Hilbert space since this is not suitable for scaling up to larger systems and higher $\U(1)$ truncations.

Our results indicate that although the topological phase transitions discussed so far do not occur in the ground state with $(\cW_x, \cW_y) = (-1, -1)$ at finite volume, in the thermodynamic limit $L\rightarrow\infty$, the topological phase transitions indeed occur in the ground state. This is completely consistent with our expectations from the lattice Lagrangian formulation, which indicate that the topological transitions must take place in the infrared in the infinite volume limit.

\subsection{The real-space diagonalization algorithm}
We utilize a real-space diagonalization algorithm to diagonalize the fermionic Hamiltonian $H_f$ coupled to gauge fields. We access different gauge configurations by using the correspondence between non-trivial Wilson loop values and twisted boundary conditions. In other words, the products $(\prod_x \sfU_x(x, 0), \prod_y \sfU_x(y, 0)) = (e^{i\theta_x}, e^{i\theta_y})$ may be accumulated and assigned to the last link as a twisted boundary condition. In other words, the $N^2$ weak-coupling superselection sectors are in one-to-one correspondence with the free Wilson fermion with $N^2$ different possible twisted boundary conditions. 

In other words, to compute the energy level of the Hamiltonian $H_f+H_m-\frac{1}{e^2}H_B$ with $(\cW_x,\cW_y) = (e^{\frac{2\pi i t_x}{N}}, e^{\frac{2\pi i t_y}{N}})$, we simply calculate the ground state of the free Wilson fermion with twisted boundary conditions $(\theta_x, \theta_y) = (\frac{2\pi t_x}{N}, \frac{2\pi t_y}{N})$, where $(t_x, t_y)\in \integers_N^x\times\integers_N^y$. We compute the half-filled ground-state energy of a free lattice Wilson fermion Hamiltonian on an $L\times L$ torus with discrete twisted boundary conditions.  The single-particle Hilbert space has dimension $2L^2$, and we assemble
\begin{equation}
H(\theta_x,\theta_y;m)=H_f(\theta_x,\theta_y)+H_m(m)
\end{equation}
where $\dim H = 2L^2$. The twists are discretized by
\begin{equation}
\theta_x=\frac{2\pi t_x}{N},\qquad \theta_y=\frac{2\pi t_y}{N},
\qquad t_x,t_y\in\{0,1,\dots,N-1\}.
\end{equation}
We construct sparse real-space hopping operators in the $x$ and $y$ directions as block-matrices that implement nearest-neighbor couplings and multiply only the wrap-around boundary link by the appropriate twist phase.  Concretely, one first builds the $2L\times 2L$ one-dimensional $x$-block $X_{\mathrm{block}}(\theta_x)$
\[
X_{\mathrm{block}}(\theta_x)\;=\;
\begin{pmatrix}
0      & M_x   & 0     & \cdots & 0 \\
0      & 0     & M_x   & \cdots & 0 \\
\vdots &       & \ddots& \ddots & \vdots \\
0      & \cdots& 0     & 0      & M_x \\
e^{i\theta_x} M_x & 0 & \cdots & 0 & 0
\end{pmatrix},
\]
and then lifts it to the full lattice by $X_{\mathrm{full}}(\theta_x)=\mathbb{1}_L\otimes X_{\mathrm{block}}(\theta_x)$. Similarly, one builds the full $y$-coupling $Y_{\mathrm{full}}(\theta_y)$ in terms of $M_y$. We then define $H_c(\theta_x,\theta_y)=X_{\mathrm{full}}({\theta_x})+Y_{\mathrm{full}}({\theta_y})$, which gives
\begin{align}
H_f(\theta_x,\theta_y)&=\frac{H_c(\theta_x,\theta_y)+H_c(\theta_x,\theta_y)^\dagger}{2},
\end{align}
so that $H_f$ is Hermitian by construction. The onsite mass term is diagonal in real space:
\begin{equation}
H_m(M)=M\,\mathbb{1}_{L^2}\otimes\sigma_3 .
\end{equation}
For each $(\theta_x,\theta_y)$ and each mass value $m$ in a scan, we diagonalize $H(\theta_x,\theta_y;m)$ to obtain eigenvalues $\{\varepsilon_j\}_{j=1}^{2L^2}$, sort them as $\varepsilon_{1}\le\cdots\le \varepsilon_{2L^2}$, and compute the half-filled ground-state energy
\begin{equation}
E_{\mathrm{hf}}(\theta_x,\theta_y;m)=\sum_{j=1}^{L^2}\varepsilon_{j} .
\end{equation}
Repeating over all $(\theta_x,\theta_y)\in\integers_N^2$ yields $N^2$ twisted sector energies $E_{\mathrm{hf}}(\theta_x,\theta_y;m)$. The same idea may be applied to excited states.
\begin{figure}[t]
    \centering
    \includegraphics[width=\linewidth]{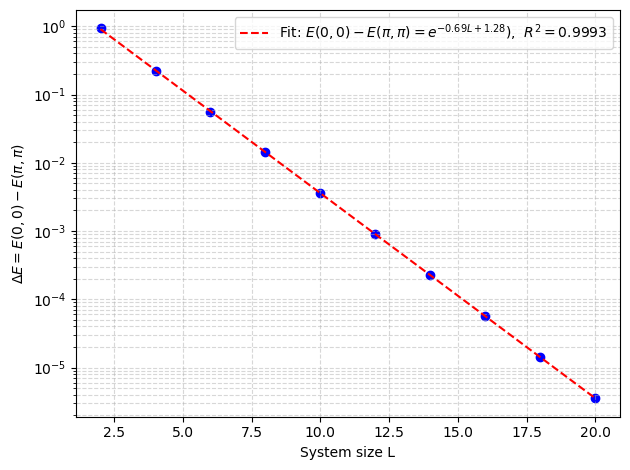}
    \caption{The exponentially vanishing degeneracy splitting between the lowest $N^2$ levels.}
    \label{fig:Deg}
\end{figure}

The complexity of each such diagonalization for a fixed mass $M$ is $O((L^2)^3N^2)$, which is (poly-)exponentially faster compared to the ED algorithm, whose complexity is at least $\Omega((2^{2L^2}N^{2L^2})^3)$. Nonetheless, the ED algorithm is instructive as it serves as a useful baseline for the analysis of potential quantum simulations.

\subsection{Results}

In Fig.~\ref{fig:FiniteSizeEffectsSpectrum}, we plot the lowest $N^2+1$ energy levels of the $\integers_N$ theory with $N=8$ for both the $2\times 2$ and $4\times 4$ cases. We note that we simply drop the magnetic contribution $H_B$ since it is simply a constant in the weakly-coupled limit. We note that the lowest level for any $N$ is always the one that has $(\cW_x, \cW_y)=(-1, -1)$ and the $(N^2)^\th$ one has $(\cW_x, \cW_y) = (1, 1)$. The left panel of Fig.~\ref{fig:FiniteSizeEffectsSpectrum} shows that although the number of levels between the $(-1, -1)$ and $(1, 1)$ approaches infinity, the gap $E_{\rm hf}(0, 0) - E_{\rm hf}(\pi, \pi)$ does \textit{not} scale with $N$! 

Interestingly, we find that the splitting between the $N^2$ twisted sectors is a finite-size effect, as suggested by the right panel of Fig.~\ref{fig:FiniteSizeEffectsSpectrum}. This idea is further confirmed by the analysis in Fig.~\ref{fig:Deg}, which shows that the gap between the $N^2$ levels is exponentially suppressed in the infinite-volume limit $L\rightarrow\infty$. This is typical of degeneracies that are lifted in finite volume due to tunneling effects.
\begin{figure}[t]
    \centering
    \includegraphics[width=\linewidth]{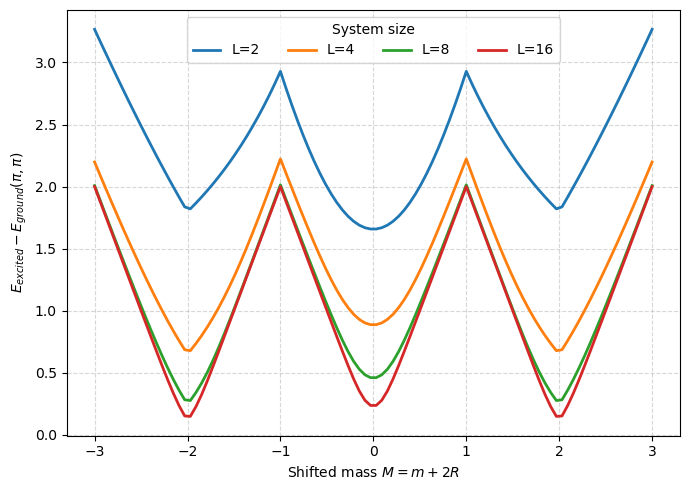}
    \caption{The gap between the true ground state with $(\cW_x, \cW_y)=(-1, -1)$ and the would-be first excited state in the infinite-volume limit for various system sizes $L$.}
    \label{fig:GapVsMforL}
\end{figure}
\begin{figure}[b]
    \centering
    \includegraphics[width=\linewidth]{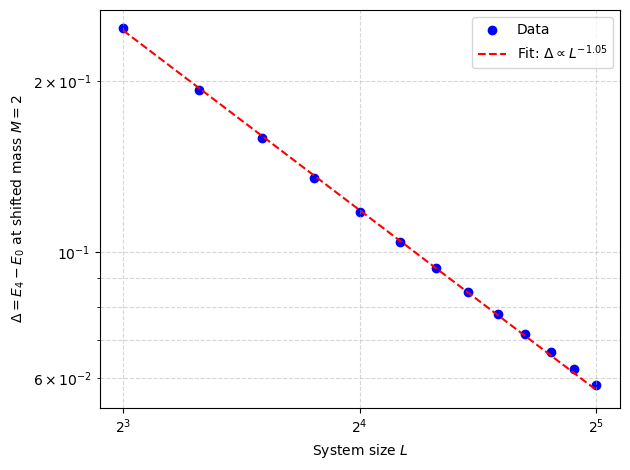}
    \caption{The power law vanishing of the gap as $L\rightarrow\infty$.}
    \label{fig:placeholder}
\end{figure}
Consistent with the idea that the infinite degeneracy is restored in the infinite-volume limit, we find that the gap between the ground state with $(\cW_x, \cW_y)=(-1, -1)$ and the would-be first excited state in the thermodynamic limit approaches zero, precisely at at $M = 0, \pm 2$, which is shown in Fig.~\ref{fig:GapVsMforL} for various system sizes. As the system size is increased, the gap at $M = 0, \pm2$ tends to zero. This is further verified by the log-log fit between the gap $\Delta$ and $L$, where we find that $\Delta(L)\sim \frac{1}{L}\rightarrow0$ in the thermodynamic limit. It is interesting to note that the scaling of the gap to zero is \textit{manifestly} different (i.e. a power law) than that of the gap between the lowest $N^2$ levels, which become degenerate exponentially rather than as a power-law. This is highly suggestive of the idea that the latter is a consequence of finite-volume tunneling effects, whilst the latter is a signature of a physical topological phase transition.

\section{Conclusion and outlook}
\label{sec:Conclusion}
Expanding on the letter \cite{Bharadwaj:2025idp}, we have studied the emergence of infrared topological phases in Hamiltonian lattice QED$_3$ with a truncated gauge group $\integers_N$ and $N_f=1,2$ flavors of Wilson fermions, motivated by near-term quantum simulations of $(2+1)$D lattice gauge theories. A key observation at weak gauge coupling is an approximate $\integers_N^x \times \integers_N^y$ symmetry generated by Wilson loops. Exploiting this structure, we gauge-fix the theory so that in the weak-coupling regime it reduces to a product of $N_f$ effectively free Wilson-fermion Hamiltonians. This limiting description supports a rich set of gapped topological phases, including integer quantum Hall (IQH) and quantum spin Hall (QSH) regimes.

By continuity, we argued that these phases persist in the fully dynamical gauge theory at finite, nonzero gauge coupling provided the coupling remains small compared to the fermion gap. Using exact diagonalization on small lattices, we verified that the ground states considered are compatible with Gauss' law and mapped out the phase structure at half-filling for both one and two flavors, extending also to finite density. This reveals metal--insulator behavior and topological phases such as Chern insulators and QSH phases. As the gauge coupling is increased, our numerics indicate that the system undergoes a transition into a strongly coupled, non-topological phase, where instanton-like effects (which are suppressed at weak coupling) become important.

More broadly, our results clarify why Wilson fermions naturally accommodate topological phases in the Hamiltonian framework, in agreement with earlier Lagrangian analyses, and emphasize that staggered-fermion formulations are not compatible with realizing such Chern phases in the same way.

Several directions are motivated by the results of this work. First, it will be important to explore the strong-coupling regime in detail and determine the structure of the resulting non-topological phase, including the role of instanton-like corrections and the possible appearance of symmetry breaking or confinement-like phenomena. A closely related challenge is to locate and classify the phase transitions separating the weak-coupling topological phases, since these critical points are generically gapless and strongly coupled, making analytic control difficult. More broadly, a comprehensive mapping of the phase diagram as a function of gauge coupling, Wilson masses, and chemical potential would be valuable, particularly at finite density where strong correlations and sign problems can obstruct classical approaches~\cite{troyer2005computational}. In this context it is also natural to investigate analogs of the Aoki phases known in Lagrangian lattice QCD$_4$ with Wilson fermions, which organize the gauge-coupling--mass plane and delineate phase boundaries at finite coupling~\cite{Aoki:1983qi,Aoki:1995ft,Sharpe:1998xm,Magnifico:2018wek,Magnifico:2019ulp}, and to ask whether similar structures arise in lattice QED$_3$ and in finite-density settings. Finally, these goals motivate the development of quantum simulation methods and quantum algorithms for detecting and mapping phase transitions, for instance using variational quantum eigensolvers and quantum phase estimation, together with measurement protocols tailored to truncated gauge-field encodings and benchmarked against exact diagonalization on small systems.

Overall, our results provide a controlled weak-coupling route to topological phases in a truncated gauge-theory setting, establish their persistence into finite coupling, and set the stage for quantum simulations to probe the strongly coupled landscape and accurately locate phase transitions beyond the reach of current classical approaches.

\textbf{Acknowledgements.} The authors would like to gratefully acknowledge conversations with Stefan Kühn, Cristina Diamantini, Pranay Naredi, Syed Muhammad Ali Hassan, Srimoyee Sen, Carsten Urbach, Zhuo Chen, Penghao Zhu, Yugo Onishi, Taige Wang, Eric D'Hoker, Theodore Jacobson and Xiaoliang Qi. The authors gratefully acknowledge the granted access to the Marvin cluster hosted by the University of Bonn. SB is supported by the Mani L. Bhaumik Institute for Theoretical Physics. This project was funded by the Deutsche Forschungsgemeinschaft (DFG, German Research Foundation) as part of the CRC 1639 NuMeriQS -- project no.\ 511713970 and under Germany's Excellence Strategy – Cluster of Excellence Matter and Light for Quantum Computing (ML4Q) EXC 2004/1 – 390534769. This work is supported with funds from the Ministry of Science, Research and Culture of the State of Brandenburg within the Centre for Quantum Technologies and Applications (CQTA). 
\begin{center}
    \includegraphics[width = 0.08\textwidth]{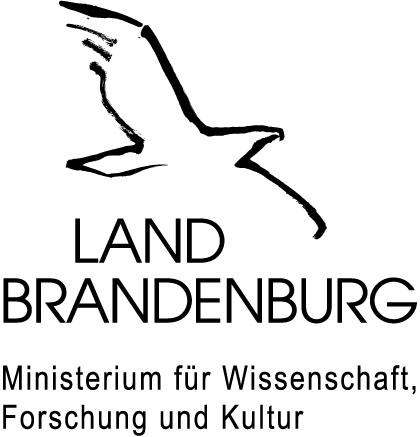}
\end{center}
This work is funded by the European Union’s Horizon Europe Frame-work Programme (HORIZON) under the ERA Chair scheme with grant agreement no. 101087126. This work is part of the Quantum Computing for High-Energy Physics (QC4HEP) working group.

\bibliographystyle{apsrev4-1}
\bibliography{main_PRD}

@article{NiuThoulessWu1985,
  title        = {Quantized Hall conductance as a topological invariant},
  author       = {Niu, Qian and Thouless, D. J. and Wu, Yong-Shi},
  journal      = {Physical Review B},
  volume       = {31},
  number       = {6},
  pages        = {3372--3377},
  year         = {1985},
  doi          = {10.1103/PhysRevB.31.3372}
}

@misc{TongGaugeTheory2018,
  author       = {Tong, David},
  title        = {Gauge Theory},
  year         = {2018},
  howpublished = {Lecture notes, Department of Applied Mathematics and Theoretical Physics, University of Cambridge},
  url          = {https://www.damtp.cam.ac.uk/user/tong/gaugetheory/gt.pdf},
  note         = {Preprint typeset in JHEP style (hyper version). Accessed: 2026-01-31}
}

@article{AvronSeilerSimon1985,
  title        = {Quantization of the Hall Conductance for General, Multiparticle Schr{\"o}dinger Hamiltonians},
  author       = {Avron, J. E. and Seiler, R. and Simon, B.},
  journal      = {Physical Review Letters},
  volume       = {54},
  number       = {4},
  pages        = {259--262},
  year         = {1985},
  doi          = {10.1103/PhysRevLett.54.259}
}

@article{AvronSeilerSimon1983,
  title        = {Homotopy and Quantization in Condensed Matter Physics},
  author       = {Avron, J. E. and Seiler, R. and Simon, B.},
  journal      = {Physical Review Letters},
  volume       = {51},
  number       = {1},
  pages        = {51--53},
  year         = {1983},
  doi          = {10.1103/PhysRevLett.51.51}
}

@article{FukuiHatsugaiSuzuki2005,
  title        = {Chern Numbers in Discretized Brillouin Zone: Efficient Method of Computing (Spin) Hall Conductances},
  author       = {Fukui, Takahiro and Hatsugai, Yasuhiro and Suzuki, Hiroshi},
  journal      = {Journal of the Physical Society of Japan},
  volume       = {74},
  number       = {6},
  pages        = {1674--1677},
  year         = {2005},
  doi          = {10.1143/JPSJ.74.1674},
  eprint       = {cond-mat/0503172},
  archivePrefix= {arXiv},
  primaryClass = {cond-mat.mes-hall}
}

@article{HastingsWen2005,
  title        = {Quasiadiabatic continuation of quantum states: The stability of topological ground-state degeneracy and emergent gauge invariance},
  author       = {Hastings, Matthew B. and Wen, Xiao-Gang},
  journal      = {Phys. Rev. B},
  volume       = {72},
  pages        = {045141},
  year         = {2005},
  doi          = {10.1103/PhysRevB.72.045141},
  eprint       = {cond-mat/0503554},
  archivePrefix= {arXiv}
}

@article{HastingsMichalakis2015,
  title        = {Quantization of Hall conductance for interacting electrons on a torus},
  author       = {Hastings, Matthew B. and Michalakis, Spyridon},
  journal      = {Communications in Mathematical Physics},
  volume       = {334},
  number       = {1},
  pages        = {433--471},
  year         = {2015},
  doi          = {10.1007/s00220-014-2167-x},
  eprint       = {1306.1258},
  archivePrefix= {arXiv},
  primaryClass = {quant-ph}
}

@article{Bharadwaj:2025idp,
    author = "Bharadwaj, Sriram and Rosanowski, Emil and Singh, Simran and Di Tucci, Alice and Peng, Changnan and Jansen, Karl and Funcke, Lena and Luo, Di",
    eprint = "2504.21828",
    archivePrefix = "arXiv",
    primaryClass = "hep-lat",
    month = "4",
    year = "2025"
}

@misc{SM,
  note = {See Supplemental Material at [URL will be inserted by publisher] for details of calculations, derivations, and additional figures}}

@article{Meth:2023wzd,
    author = "Meth, Michael and others",
    title = "{Simulating two-dimensional lattice gauge theories on a qudit quantum computer}",
    eprint = "2310.12110",
    archivePrefix = "arXiv",
    primaryClass = "quant-ph",
    reportNumber = "To be published in Nature Physics (2025)",
    doi = "10.1038/s41567-025-02797-w",
    journal = "Nature Phys.",
    volume = "21",
    number = "4",
    pages = "570--576",
    year = "2025"
}

@article{Cochran:2024rwe,
    author = "Cochran, Tyler A. and others",
    title = "{Visualizing Dynamics of Charges and Strings in (2+1)D Lattice Gauge Theories}",
    eprint = "2409.17142",
    archivePrefix = "arXiv",
    primaryClass = "quant-ph",
    month = "9",
    year = "2024"
}

@article{chen2022simulating,
  title={Simulating 2+1d lattice quantum electrodynamics at finite density with neural flow wavefunctions},
  author={Chen, Zhuo and Luo, Di and Hu, Kaiwen and Clark, Bryan K},
  journal={arXiv:2212.06835},
  year={2022},
  doi = {10.48550/arXiv.2212.06835}
}

@article{Sen:2020srn,
    author = "Sen, Srimoyee",
    title = "{Chern insulator transitions with Wilson fermions on a hyperrectangular lattice}",
    eprint = "2008.01743",
    archivePrefix = "arXiv",
    primaryClass = "hep-th",
    doi = "10.1103/PhysRevD.102.094520",
    journal = "Phys. Rev. D",
    volume = "102",
    number = "9",
    pages = "094520",
    year = "2020"
}

@article{Golterman:1992ub,
    author = "Golterman, Maarten F. L. and Jansen, Karl and Kaplan, David B.",
    title = "{Chern-Simons currents and chiral fermions on the lattice}",
    eprint = "hep-lat/9209003",
    archivePrefix = "arXiv",
    reportNumber = "UCSD-PTH-92-28, WASH-U-HEP-92-62",
    doi = "10.1016/0370-2693(93)90692-B",
    journal = "Phys. Lett. B",
    volume = "301",
    pages = "219--223",
    year = "1993"
}

@article{Qi:2006xub,
    author = "Qi, Xiao-Liang and Wu, Yong-Shi and Zhang, Shou-Cheng",
    title = "{Topological quantization of the spin Hall effect in two-dimensional paramagnetic semiconductors}",
    doi = "10.1103/PhysRevB.74.085308",
    journal = "Phys. Rev. B",
    volume = "74",
    number = "8",
    pages = "085308",
    year = "2006"
}

@article{Bernevig:2006uvn,
    author = "Bernevig, B. Andrei and Hughes, Taylor L. and Zhang, Shou-Cheng",
    title = "{Quantum Spin Hall Effect and Topological Phase Transition in HgTe Quantum Wells}",
    doi = "10.1126/science.1133734",
    journal = "Science",
    volume = "314",
    number = "5806",
    pages = "1133734",
    year = "2006"
}

@article{luo2020framework,
  title={Framework for simulating gauge theories with dipolar spin systems},
  author={Luo, Di and Shen, Jiayu and Highman, Michael and Clark, Bryan K and DeMarco, Brian and El-Khadra, Aida X and Gadway, Bryce},
  journal={Physical Review A},
  volume={102},
  number={3},
  pages={032617},
  year={2020},
  publisher={APS}
}

@article{gonzalez2024observation,
  title={Observation of string breaking on a (2+ 1) D Rydberg quantum simulator},
  author={Gonzalez-Cuadra, Daniel and Hamdan, Majd and Zache, Torsten V and Braverman, Boris and Kornjaca, Milan and Lukin, Alexander and Cantu, Sergio H and Liu, Fangli and Wang, Sheng-Tao and Keesling, Alexander and others},
  journal={arXiv preprint arXiv:2410.16558},
  year={2024}
}

@article{sun2023quantum,
  title={Quantum Computation and Simulation using Fermion-Pair Registers},
  author={Sun, Xiangkai and Luo, Di and Choi, Soonwon},
  journal={arXiv preprint arXiv:2306.03905},
  year={2023}
}

@article{troyer2005computational,
  title={Computational Complexity and Fundamental Limitations<? format?> to Fermionic Quantum Monte Carlo Simulations},
  author={Troyer, Matthias and Wiese, Uwe-Jens},
  journal={Physical review letters},
  volume={94},
  number={17},
  pages={170201},
  year={2005},
  publisher={APS}
}

@article{Aoki:1983qi,
    author = "Aoki, Sinya",
    title = "{New Phase Structure for Lattice QCD with Wilson Fermions}",
    reportNumber = "UT-421-TOKYO",
    doi = "10.1103/PhysRevD.30.2653",
    journal = "Phys. Rev. D",
    volume = "30",
    pages = "2653",
    year = "1984"
}

@article{Aoki:1995ft,
    author = "Aoki, S.",
    editor = "Nakamura, A. and Kanaya, K. and Karsch, F.",
    title = "{On the phase structure of QCD with Wilson fermions}",
    eprint = "hep-lat/9509008",
    archivePrefix = "arXiv",
    reportNumber = "UTHEP-318",
    doi = "10.1143/PTPS.122.179",
    journal = "Prog. Theor. Phys. Suppl.",
    volume = "122",
    pages = "179--186",
    year = "1996"
}

@article{Sharpe:1998xm,
    author = "Sharpe, Stephen R. and Singleton, Jr, Robert L.",
    title = "{Spontaneous flavor and parity breaking with Wilson fermions}",
    eprint = "hep-lat/9804028",
    archivePrefix = "arXiv",
    reportNumber = "UW-PT-98-2",
    doi = "10.1103/PhysRevD.58.074501",
    journal = "Phys. Rev. D",
    volume = "58",
    pages = "074501",
    year = "1998"
}

@article{Magnifico:2019ulp,
    author = {Magnifico, G. and Vodola, D. and Ercolessi, E. and Kumar, S. P. and M\"uller, M. and Bermudez, A.},
    title = "{$\mathbb{Z}_N$ gauge theories coupled to topological fermions: QED$_2$ with a quantum-mechanical $\theta$ angle}",
    eprint = "1906.07005",
    archivePrefix = "arXiv",
    primaryClass = "cond-mat.quant-gas",
    doi = "10.1103/PhysRevB.100.115152",
    journal = "Phys. Rev. B",
    volume = "100",
    number = "11",
    pages = "115152",
    year = "2019"
}

@article{Magnifico:2018wek,
    author = {Magnifico, G. and Vodola, D. and Ercolessi, E. and Kumar, S. P. and M\"uller, M. and Bermudez, A.},
    title = "{Symmetry-protected topological phases in lattice gauge theories: topological QED$_2$}",
    eprint = "1804.10568",
    archivePrefix = "arXiv",
    primaryClass = "cond-mat.quant-gas",
    doi = "10.1103/PhysRevD.99.014503",
    journal = "Phys. Rev. D",
    volume = "99",
    number = "1",
    pages = "014503",
    year = "2019"
}

\appendix

\clearpage

\onecolumngrid
\begin{center}
	\noindent\textbf{Supplementary Material}
	\bigskip
		
	\noindent\textbf{\large{}}
\end{center}

\onecolumngrid

\renewcommand{\thefigure}{S\arabic{figure}}
\setcounter{secnumdepth}{2}

\section{Supplementary figures for staggered fermions and na\"ive Wilson fermions}\label{App:StagFig}
\begin{figure}[H]
    \centering
    \subfloat[Na\"ive Wilson fermions with $R=0$]{\includegraphics[width=0.42\linewidth]{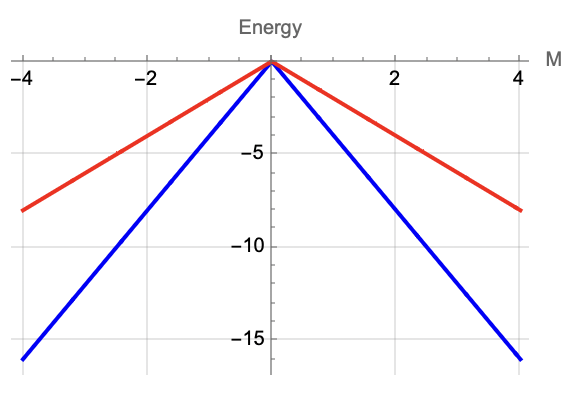}}
    \quad\quad\quad\quad
    \subfloat[Staggered fermions]{\includegraphics[width=0.431\linewidth]{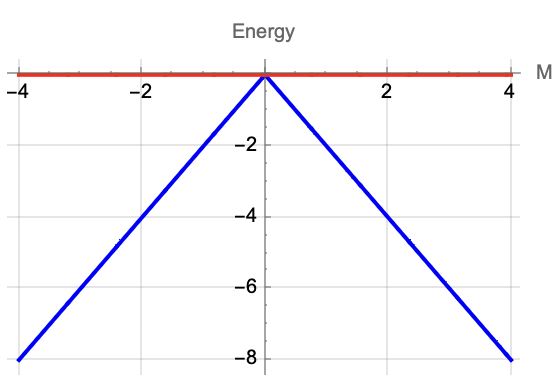}}
    \caption{ The lowest two energies for na\"ive Wilson and Staggered fermions: note the absence of transitions at $M=\pm 2$.}
    \label{fig:StaggeredNaive}
\end{figure}
In this subsection, we present the analog of the result in Fig.~\ref{fig:LevelCrossing} for the case of staggered fermions and na\"ive Wilson fermions with $R = 0$ (see Fig.~\ref{fig:StaggeredNaive}). These results were produced by exact diagonalization on a $2\times 2$ lattice. Unlike the case of Wilson fermions with $R = 1$, we find that there is only a single gapless point at $M=0$ (which is metallic), and no topological transitions or corresponding level crossings at $M = \pm 2$. Away from $M=0$, the theory is in a trivially insulating phase.
\section{Order of metal-insulator transition}\label{Order}
\begin{figure*}[h]
    \centering
    \includegraphics[width=0.9\linewidth]{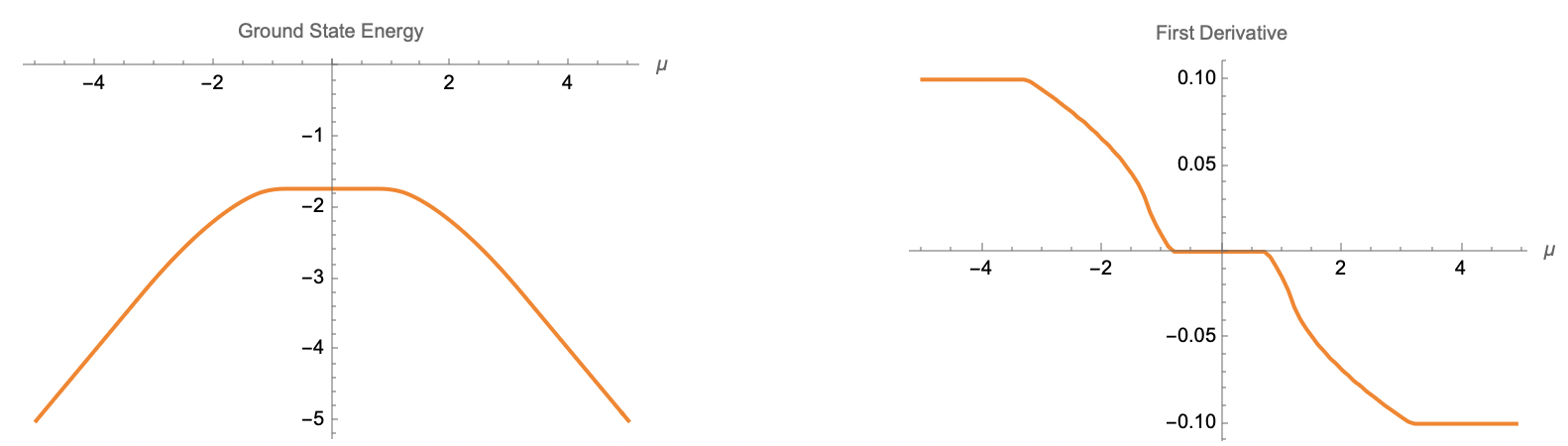}
    \caption{A plot of the ground-state energy and its first derivative at $M = -1.2$ on a $32\times 32$ spatial lattice. Clearly, the energy is smooth at the transition points $\mu = \pm 0.8, \pm 3.2$, while its first derivative $\frac{dE}{d\mu}$ is non-analytic at those points, indicating a phase transition.}
    \label{fig:Transition}
\end{figure*}

To analyze the order of the metal-insulator transitions, we examine the ground-state energy and its first derivative as a function of $\mu$ (see Fig.~\ref{fig:Transition}).
\newpage
\section{Supplementary figures for the $N_f=2$ theory at finite density with equal masses}\label{App:SupplementaryFigs}
\subsection{Structure of the four-band model for $M_1=M_2=M$}
When $\mu = 0$, the lowest energy band is highly degenerate $\ket{z_1, z_2}\equiv z_1\ket{\uparrow, u_-}+ z_2\ket{\downarrow, u_-}$, where $\abs{z_1}^2+\abs{z_2}^2 = 1\text{ and } z_1, z_2\in\complex$. Clearly, $(z_1, z_2)$ parametrize points on ${S}^3\cong_\text{manifold}\SU(2)_F$, which is consistent with the existence of an $\SU(2)_F$ flavor symmetry. As the chemical potential is turned on, the degenerate moduli-space of vacua is ``lifted", and we have four bands (see Fig.~\ref{fig:VacuumStructure} in Supplementary Material \ref{App:SupplementaryFigs}).
\begin{figure}[h]
        \centering
        \subfloat[$\mu = -3.5$ \label{fig:a}]{\includegraphics[width=0.4\textwidth]{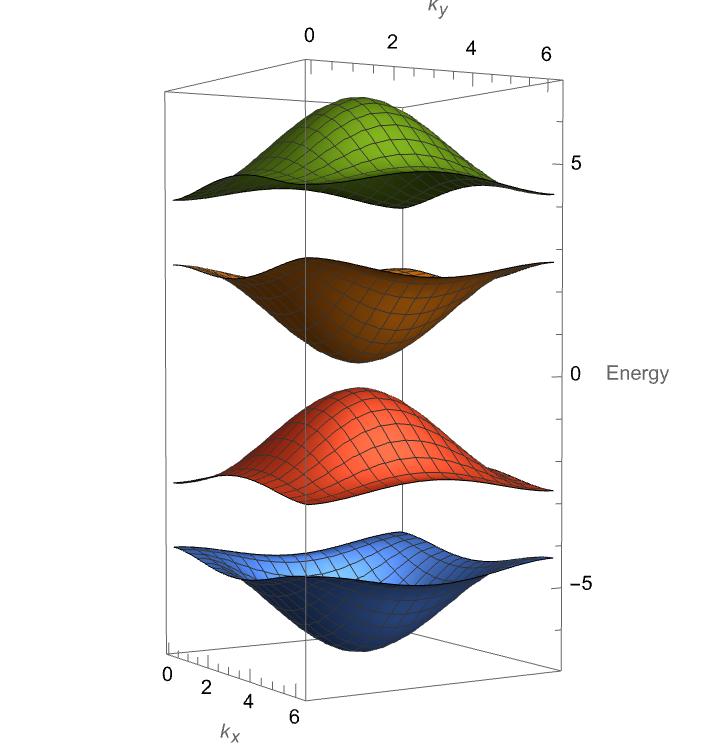}}
        \hfill
        \subfloat[$\mu = -1.5$ \label{fig:b}]{\includegraphics[width=0.4\textwidth]{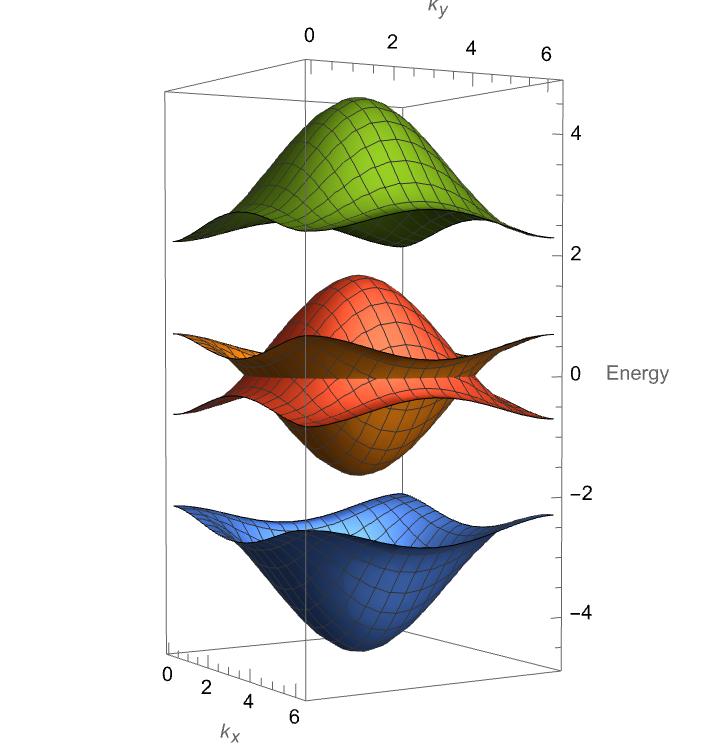}}
        \hfill
        \subfloat[$\mu = 1.5$ \label{fig:c}]{\includegraphics[width=0.4\textwidth]{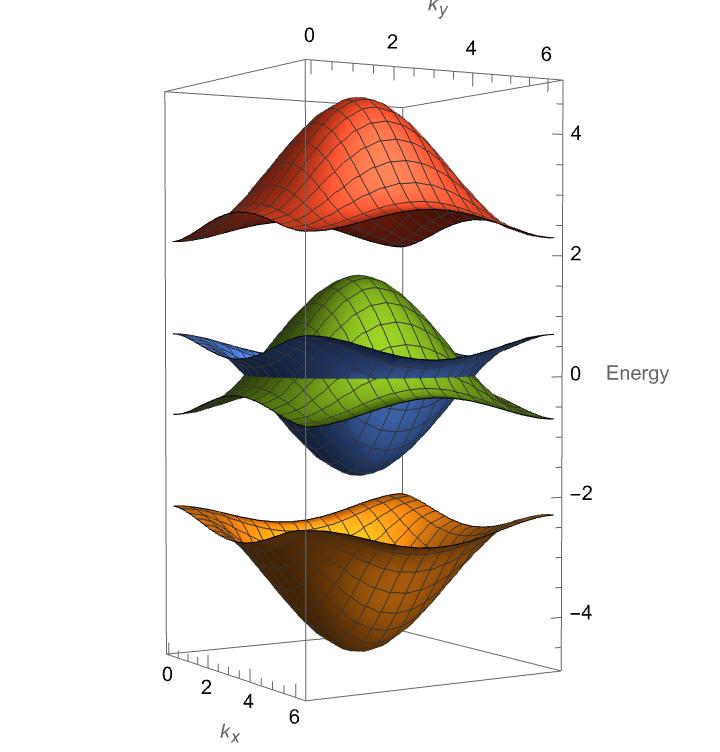}}
        \hfill
        \subfloat[$\mu = 3.5$ \label{fig:d}]{\includegraphics[width=0.4\textwidth]{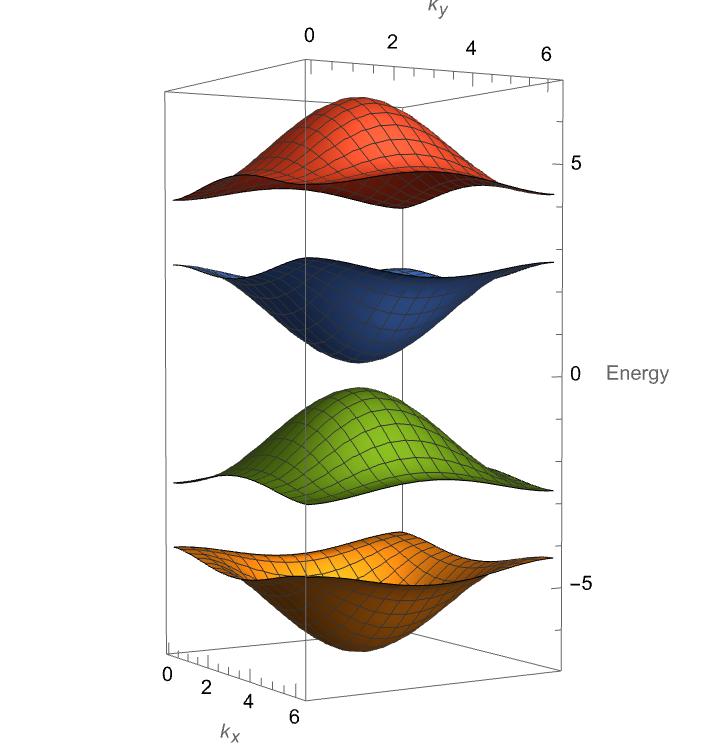}}
        \caption[]
        {\small The above figures show a plot of the four bands over momentum-space for various values of the relative chemical potential $\mu$ and fixed mass $M = 0.8$. The blue ($-$) and red ($+$) bands comprise the states $\ket{\downarrow, u_\pm(k)}$ respectively, while the orange ($-$) and green ($+$) ones comprise the energy levels $\ket{\uparrow, u_\pm(k)}$ respectively. Tuning $\mu$ drives the theory into qualitatively distinct phases. At half-filling, the theory is gapped in the regimes of small and large $\mu$, as shown in \ref{fig:a} and \ref{fig:d}. On the other hand, the theory at half-filling is gapless and displays metallic behavior in the crossover regions in \ref{fig:b} and \ref{fig:c}.} 
        \label{fig:VacuumStructure}
    \end{figure}

\subsection{Gaplessness at $\mu = 0$ when $M_1=M_2=M = 0, \pm 2$}

\begin{figure}[h]
        \centering
        \subfloat[$M = 0$\label{fig:a}]{\includegraphics[width=0.34\textwidth]{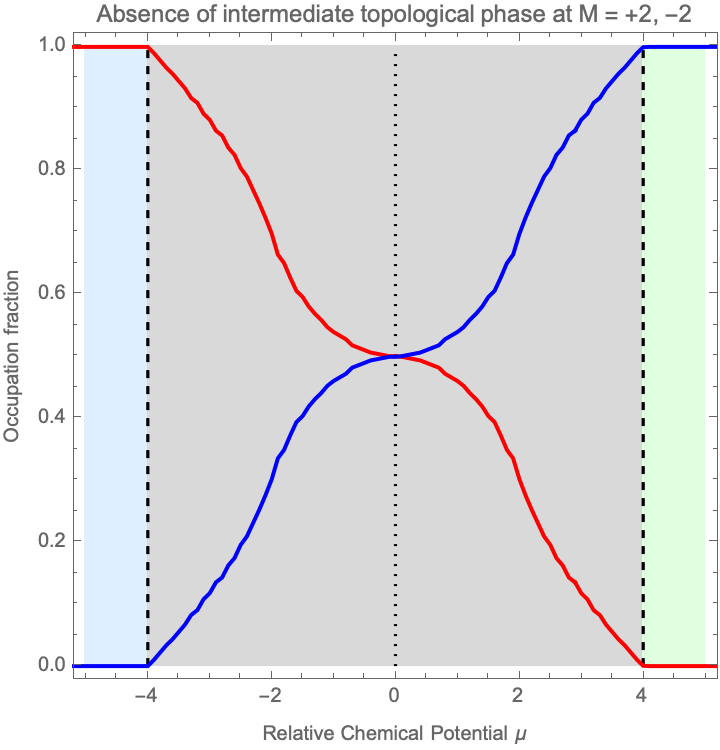}}
        \subfloat[$M = \pm 2$\label{fig:b}]{\includegraphics[width=0.6\textwidth]{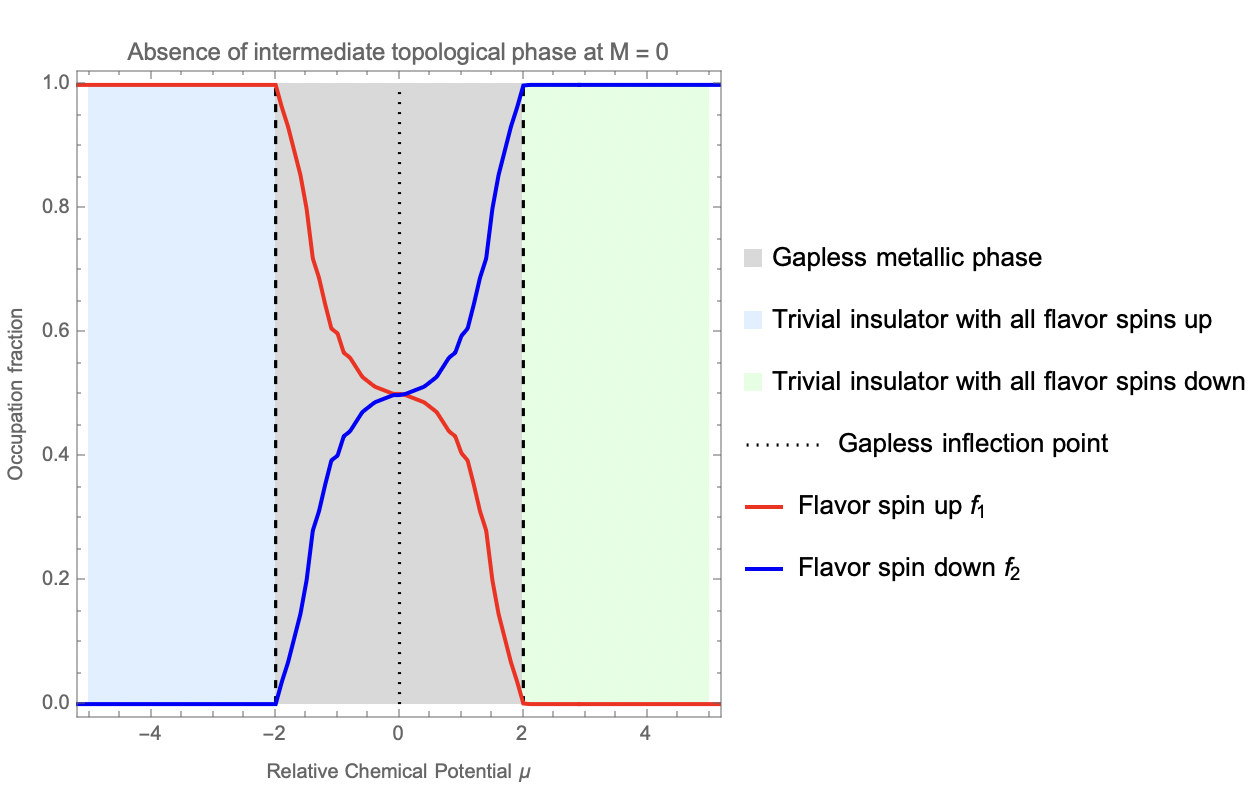}}
        \caption[]
        {\small The average occupation fraction in the vacuum $f_a$ versus the relative chemical potential $\mu$. Interestingly, we find that the yellow gapped regime of Fig.~\ref{fig:OccNo} collapses to form a gapless inflection point. This is crucial for consistency with the topological phase diagram at zero density since we know the theory to be gapless at $M = 0, \pm 2$ when $\mu=0$. These calculations were performed on a $32\times 32$ spatial lattice.} 
        \label{fig:M0pm2}
    \end{figure}

We know that the theory is gapless at $M=-2,0,+2$ when $\mu = 0$, which means that the intermediate yellow region in Fig.~\ref{fig:OccNo} cannot exist for those values of $M$. Indeed, this is confirmed explicitly by the plots in Fig.~\ref{fig:M0pm2}.

\newpage
\subsection{Occupation fraction versus singlet mass $M$ for various fixed $\mu$}
\begin{figure}[h]
        \centering
        \subfloat[$\mu = 0.8$: four distinct gapped phases\label{fig:Ma}]{\includegraphics[width=0.4\textwidth]{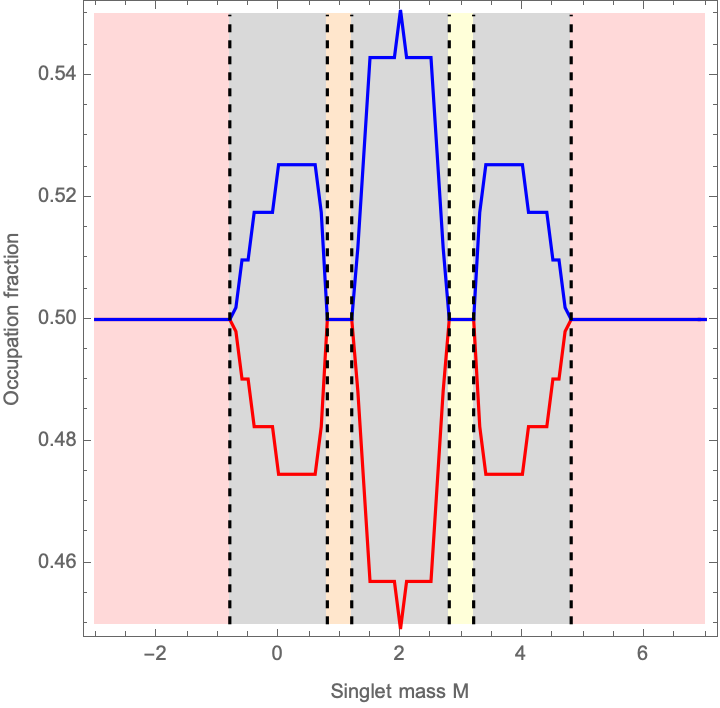}}  
        \hfill
        \subfloat[$\mu = 1.5$: two distinct gapped phases\label{fig:Mb}]{\includegraphics[width=0.4\textwidth]{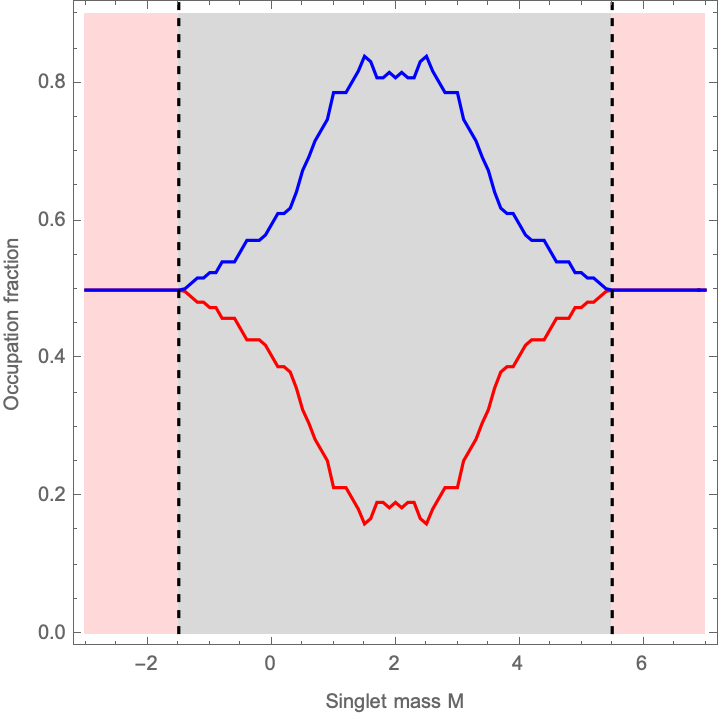}}
        
        \bigskip
        
        \subfloat[$\mu = 2.5$: three distinct gapped phases\label{fig:Mc}]{\includegraphics[width=0.4\textwidth]{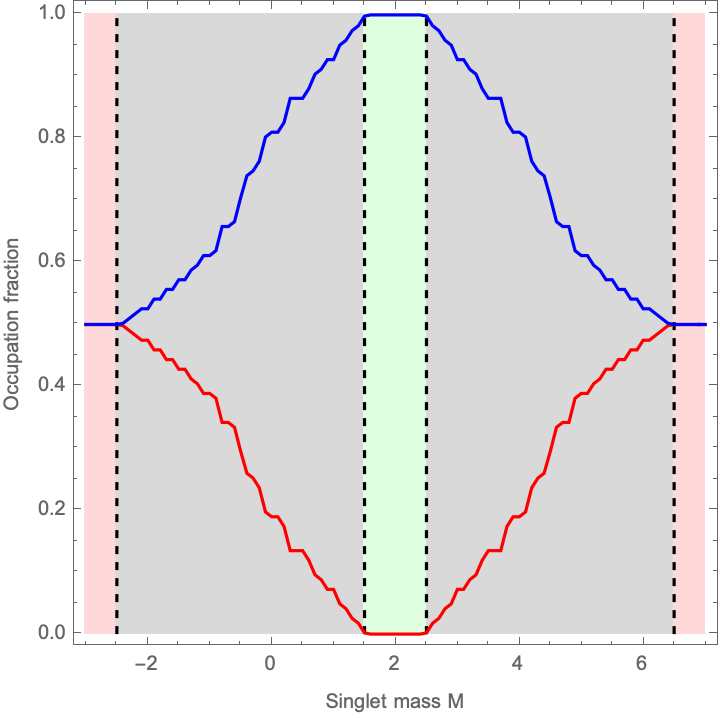}}
        \hfill
        \subfloat[\small Legend for the $f_a$ vs $M$ plots\label{fig:Md}]{\includegraphics[width=0.4\textwidth]{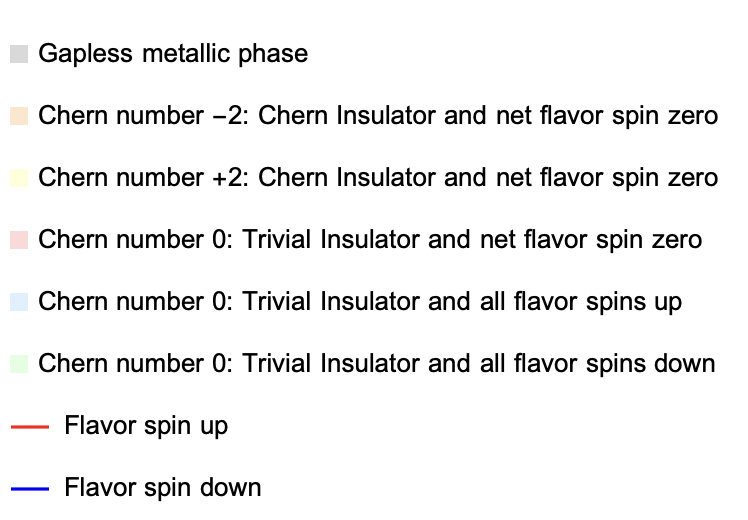}}
        \caption[]
        {\small The average occupation fraction $f_a$ in the vacuum versus the singlet mass $M$ for various fixed values of the chemical potential $\mu$. We observe that there are three distinct qualitative regimes: small $\mu\ll 1$ with four distinct gapped phases (figure \ref{fig:Ma}), intermediate $\mu\sim 1$ with two gapped phases (figure \ref{fig:Mb}), and large $\mu\gg 1$ with three distinct gapped phases (figure \ref{fig:Mc}). Each of the colored gapped phases are distinguished by their vacuum Chern number and net flavor spin, as indicated by the legend in Fig.~\ref{fig:Md}. The above calculations were performed on a $16\times 16$ spatial lattice.} 
        \label{fig:faVsM}
    \end{figure}

We also studied the occupation fractions $f_a$ as function of the singlet mass $M$ for fixed values of the chemical potential $\mu$. This analysis reveals (see Fig.~\ref{fig:faVsM}) the existence of three qualitatively distinct regimes. When $\mu\ll 1$, we find that the theory has four distinct gapped phases as we dial $M$ that are distinguished by their vacuum Chern numbers and quantum numbers under the flavor symmetry. When $\mu\sim 1$, then we find that the theory can only have two distinct gapped phases as we dial $M$. And, lastly, when $\mu \gg 1$, the theory will have three distinct gapped phases as we dial $M$. 

\newpage
\section{Consistency with the phase diagram at zero density}
\begin{figure}[H]
    \centering
    \includegraphics[width=0.7\linewidth, scale=1]{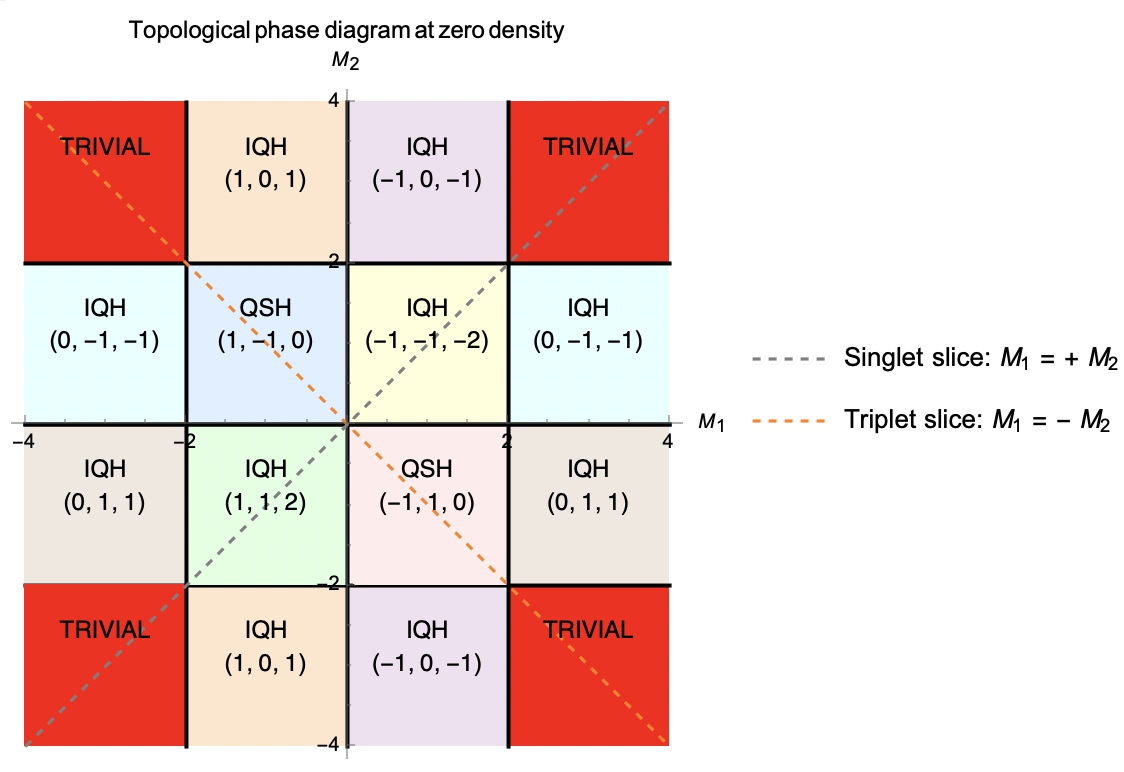}
    \caption{The above plot shows the topological phase diagram for the most general diagonal mass matrix with equal Wilson couplings $R=1$. As the legend indicates, the theory exhibits a variety of gapped phases: the IQH, QSH, and trivial insulator phases. The dashed blue line $M_1 = M_2$ is highlighted so one can easily verify consistency with the results found at finite density. The same holds for the line with $M_1 = -M_2$, which is the dashed orange line. 
    \label{fig:ZeroDensityNf=2}}
\end{figure}

At zero density, it is straightforward to map out the topological phase diagram as a function of the masses $M_{a=1,2} = m_a + 2$ of the two Dirac fermions respectively. We have summarized the results of this analysis in Fig.~\ref{fig:ZeroDensityNf=2}, which shows a rich phase diagram. The black lines denote slices at which the lattice theory is gapless, and across which there are phase transitions. In the gapped regimes, the theory exhibits IQHE, QSHE, or is a trivial insulator. 

\end{document}